\documentclass[aps,prl,twocolumn,superscriptaddress,nofootinbib,longbibliography]{revtex4-2}
\usepackage{amsmath,amssymb,bm,mathtools}
\usepackage{xcolor,graphicx}
\usepackage[colorlinks=true,linkcolor=blue!55!black,citecolor=blue!55!black,urlcolor=blue!55!black]{hyperref}

\newcommand{\C}{\mathbb{C}}

\newcommand{\dd}{\mathrm{d}}
\newcommand{\e}{\mathrm{e}}
\newcommand{\im}{\mathrm{i}}
\newcommand{\Tr}{\operatorname{Tr}}

\newcommand{\rad}[1]{\mathcal{R}_{#1}}
\newcommand{\gauged}[1]{\mathcal{S}_{#1}}
\newcommand{\AId}{\mathrm{AI}^{\dagger}}
\newcommand{\AIId}{\mathrm{AII}^{\dagger}}
\newcommand{\pnn}{p_{\mathrm{NN}}}
\newcommand{\pnnN}[1]{p_{\mathrm{NN};(#1)}}

\makeatletter
\newcommand{\equalcontrib}{\frontmatter@footnote{These two authors contributed equally to this work.}}
\makeatother

\begin{document}

\title{Exact joint eigenvalue densities of non-Hermitian random matrices are Calogero scattering states}

\author{Zhenyu Xiao\equalcontrib}
\email{zyxiao@princeton.edu}
\affiliation{Princeton Quantum Initiative, Princeton University, Princeton, New Jersey 08544, USA}
\affiliation{Princeton Center for Theoretical Science, Princeton University, Princeton, New Jersey 08544, USA}
\author{Ze Chen\equalcontrib}
\affiliation{Department of Physics, Princeton University, Princeton, New Jersey 08544, USA}
\affiliation{Department of Electrical and Computer Engineering, Princeton University, Princeton, New Jersey 08544, USA}
\author{Shinsei Ryu}
\email{shinseir@princeton.edu}
\affiliation{Department of Physics, Princeton University, Princeton, New Jersey 08544, USA}

\date{\today}

\begin{abstract}
Determining exact joint eigenvalue densities is central to random matrix theory.
We solve this long-standing problem for non-Hermitian matrices with transposition symmetry (complex symmetric and complex self-dual) at arbitrary matrix size.
Up to a Vandermonde factor, they are scattering-state wave functions of the Calogero model, a line of particles interacting through an inverse-square potential, with the coupling strength set by the symmetry.
In contrast to many previously known joint densities, the densities cannot be written as a gas of eigenvalues with pairwise interactions.
We further compute the complex level spacing distributions and two-point spectral correlation functions, which carry power-law tails, absent in a Coulomb gas.
Our results shed light on the interplay among random matrices, integrability, and symmetry.
\end{abstract}

\maketitle

\emph{Introduction.---}
Random matrix theory (RMT) is a common language for diverse fields of physics.
It began with Wigner's realization that heavy-nucleus spectra follow random-matrix statistics~\cite{Wigner1951,Wigner1958}.
Chaotic quantum systems display RMT level statistics~\cite{Bohigas1984,Haake2018}, from billiards~\cite{McDonaldKaufman1979,StoeckmannStein1990} to disordered~\cite{Altshuler1988,Shklovskii1993,Mirlin2000,EversMirlin2008} and strongly correlated systems~\cite{Poilblanc1993,OganesyanHuse2007,PalHuse2010,SerbynMoore2016,DAlessio2016,Abanin2019}.
Beyond quantum chaos, its reach extends to the zeros of the Riemann zeta function~\cite{Montgomery1973,Mehta2004}.

Non-Hermitian random matrices arise in diverse physical settings~\cite{GarciaGarcia2002,GarciaGarcia2003}, including quantum chromodynamics~\cite{Stephanov1996,Osborn2004,AkemannWettig2004,KanazawaWettig2021}, quantum transport~\cite{Haake1992,Beenakker1997,Beenakker2015}, and stochastic processes~\cite{May1972,May1976,Sommers1988,Sompolinsky1988,NelsonShnerb1998,MuruganVaikuntanathan2017}.
Recent interest in open quantum systems has made their universality more pressing still~\cite{Sa2020Liouvillian,Sa2020Kraus,Sa2022Lindbladian,Costa2023}.
Markovian density-matrix evolution is generated by a Lindbladian~\cite{Gorini1976,Lindblad1976}, which is non-Hermitian, and continuously monitored systems admit a non-Hermitian Hamiltonian representation~\cite{Dalibard1992,XiaoOhtsukiKawabata2025,XiaoKawabata2026}.
The conjecture that chaos in open quantum systems is characterized by non-Hermitian RMT level statistics has since been studied extensively~\cite{Grobe1988,Denisov2019,Akemann2019,Can2019,Sa2020,Li2021,Kulkarni2022,Shivam2023,Kawabata2023SVD,Xiao2024,GarciaGarcia2023,Li2024DSFF,Richter2025}.

Symmetry is the organizing principle of RMT.
From time-reversal symmetry (TRS), Dyson built his threefold way for Hermitian matrices, giving three universality classes of level correlations~\cite{Dyson1962c}.
Non-Hermitian symmetry is subtler.
At the very beginning of non-Hermitian RMT, Ginibre established a parallel threefold way from TRS.
His three ensembles, however, share the same bulk statistics away from the real axis~\cite{Grobe1989,Hamazaki2020,Ginibre1965,ByunForrester2024}.
The generic bulk is instead organized by the extension of time reversal defined by transposition rather than conjugation, termed time-reversal symmetry$^{\dagger}$ (TRS$^{\dagger}$)~\cite{BernardLeClair2002,Magnea2008,KawabataNatCommun2019,Kawabata2019}, which generates a threefold way of its own~\cite{Kulkarni2025,Akemann2025a,Akemann2026} (see the next section); both symmetries are widely realized in physical systems~\cite{GarciaGarcia2022,Kawabata2023,Xiao2022,ElGanainy2018,Altland2021,LuoOhtsukiShindou2021,LuoXiao2022,Sa2023Symmetry,Li2024,Wold2025,Yoshimura2024}.

The joint eigenvalue density carries the complete information about level correlations, and obtaining it is therefore one of the central goals of RMT.
For Hermitian matrices~\cite{Wigner1958,Mehta2004}, it followed soon after the symmetry classification itself~\cite{Dyson1962c,Verbaarschot1994,Zirnbauer1996,AltlandZirnbauer1997,Caselle2004}.
For non-Hermitian matrices, 
however, the joint densities have remained unknown except in special cases.
Ginibre solved two of his three ensembles~\cite{Ginibre1965,ByunForrester2024}, and the remaining real one resisted for a quarter of a century~\cite{Lehmann1991,EdelmanKostlanShub1994,ForresterNagao2007}.
The two classes that carry TRS$^{\dagger}$, complex symmetric (class $\AId$) and complex self-dual (class $\AIId$) matrices, have proved harder still~\cite{Hamazaki2020,Kulkarni2025,Akemann2025a}, and have remained unsolved beyond $2\times2$ matrices~\cite{KawabataRyu2026,LiuZhang2024,Forrester2025}, despite recent progress on one-point densities~\cite{Akemann2025b,Crumpton2026}.
The obstruction is structural: unlike in the symmetry-free class, the eigenvalues admit no description as a gas with pairwise interactions~\cite{Hastings2001}; the problem is genuinely many-body, as we show in this work.

In this Letter, we obtain the exact joint eigenvalue densities of complex symmetric and complex self-dual random matrices in the Gaussian ensembles, at arbitrary matrix size.
We achieve this by identifying an integrable structure that emerges in the joint density: that of the Calogero model, particles on a line interacting through an inverse-square potential~\cite{Calogero1971,Hallnas2024,Touzo2024,Liegeois2025}.
The densities are its asymptotically free scattering states, with the interaction strength set by the symmetry class.
The connection is through analytic continuation: each eigenvalue $z$ enters as a complex position, and $\im\bar z$ as the corresponding momentum.
The three TRS$^\dagger$ classes inherit three distinct levels of integrable structure: the Calogero model is free for Ginibre's ensemble, algebraically integrable for class $\AIId$, and only Liouville integrable for class $\AId$.
We use the result to obtain exact finite-$N$ complex level-spacing distributions and find a power-law tail in the two-point spectral correlation function.
Finally, we find both new classes realized in the reflection matrix of a Hermitian disordered conductor.

\emph{Threefold ways and exact solution.---}
Without Hermiticity, $H^{*}\neq H^{\mathrm T}$, and time reversal splits in two,
\begin{equation}
 \mathcal T_+H^{*}\mathcal T_+^{-1}=H\ \ (\mathrm{TRS}),
 \quad
 \mathcal C_+H^{\mathrm T}\mathcal C_+^{-1}=H\ \ (\mathrm{TRS}^{\dagger}),
 \label{eq:trs}
\end{equation}
with unitary $\mathcal T_+,\mathcal C_+$ obeying $\mathcal T_+\mathcal T_+^{*}=\pm1$ and $\mathcal C_+\mathcal C_+^{*}=\pm1$.
Absence of the symmetry gives class A in each threefold way, while TRS$^{\dagger}$ (TRS) with sign $+1$ and $-1$ give classes $\AId$ and $\AIId$ (AI and AII), respectively~\cite{BernardLeClair2002,Magnea2008,Kawabata2019}.
TRS makes the spectrum symmetric about the real axis: eigenvalues are either real or come in pairs $(z,z^{*})$.
In the bulk, the partner lies many mean spacings away and drops out of the local correlations, so classes AI and AII share the bulk statistics of A~\cite{Hamazaki2020,Xiao2022,ByunForrester2024}.
TRS$^{\dagger}$, by contrast, leaves every eigenvalue in place and constrains its eigenvectors instead, so it acts throughout the spectrum and reshapes the bulk statistics~\cite{Hamazaki2020,Kulkarni2025,Akemann2025a,Akemann2026}.

We take Gaussian ensembles in the three TRS$^{\dagger}$ classes: a generic complex matrix for class A, a complex symmetric matrix $H=H^{\mathrm T}$ for class $\AId$, and a complex self-dual matrix $H=JH^{\mathrm T}J^{-1}$, $J=I_N\otimes\im\sigma_y$, for class $\AIId$, each distributed as
\begin{equation}
\dd P(H)\propto\e^{-\Tr(H^\dagger H)/\kappa}\,\dd H,
\label{eq:ensembles}
\end{equation}
where $\kappa$ is the Kramers degeneracy ($\kappa=2$ in class $\AIId$, $\kappa=1$ otherwise).

\textbf{Theorem.} For class $\mathcal C=\AId$, A, or $\AIId$, set $\beta=1,2,4$, respectively, and let $\rho^{\mathcal C}$ denote the joint probability density of the $N$ distinct eigenvalues $\bm z=(z_1,\ldots,z_N)$.
Then
\begin{equation}
\rho^{\mathcal C}(\bm z)
=Z_{N,\beta}^{-1}\,|\Delta(\bm z)|^{\beta}\,
\Psi_{\beta/2}(\bm x=\bm z,\;\bm p=\im\bar{\bm z}),
\label{eq:unified}
\end{equation}
where $Z_{N,\beta}$ is a normalization constant and $\Delta(\bm z)=\prod_{i<j}(z_i-z_j)$.
Here, $\bm x=(x_1,\ldots,x_N)$, $\bm p=(p_1,\ldots,p_N)$, and $\Psi_k(\bm x,\bm p)$, with $k=\beta/2$, is the asymptotically free scattering eigenfunction of the $N$-body rational Calogero model
\begin{equation}
\mathcal H_k=\sum_i p_i^{\,2}+\sum_{i<j}\frac{2k(k-1)}{(x_i-x_j)^2},
\qquad p_i=-\im\,\partial_{x_i},
\label{eq:HCal}
\end{equation}
``asymptotically free'' meaning that it reduces to the single plane wave $\e^{\im\bm p\cdot\bm x}$ when all pair separations are large~\cite{Opdam1993,Chalykh1990,Melin2024}.
One such eigenfunction exists for every momentum vector $\bm p$, and Eq.~\eqref{eq:unified} uses the whole family: each configuration $\bm z$ is evaluated in its own member, the one with momenta $\bm p=\im\bar{\bm z}$.

The Calogero Hamiltonian is Liouville integrable for all real $k$~\cite{Olshanetsky1983} and algebraically integrable for integer $k$~\cite{Chalykh1990,Chalykh1999}, and free for $k=0,1$.
Thus classes, $\AId$, A, and $\AIId$, with $k= \beta/2 = 1/2,1,2$, realize distinct integrable structures, with analytical scattering states given below.

For class A ($\beta=2$, $k=1$), $\mathcal H_k$ is free and the scattering state is the plane wave $\Psi_1(\bm x,\bm p)=\e^{\im\bm p\cdot\bm x}$.
Equation~\eqref{eq:unified} yields the Ginibre result $\rho^{\mathrm A}(\bm z)\propto \e^{-\sum_i|z_i|^2}|\Delta(\bm z)|^2$ \cite{Ginibre1965,ByunForrester2024}, which is a pair gas: the eigenvalues interact through the two-body potential $-2\log|z_i-z_j|$ alone.

For class $\AIId$ ($\beta=4$, $k=2$), $\mathcal H_k$ is algebraically integrable,
and $\Psi_2$ is a Baker--Akhiezer function
\cite{Chalykh1990,Chalykh1999}: it is given in closed form for every $N$
by an explicit operator formula [Eq.~\eqref{eq:AIIberest}] \cite{Chalykh1999,Felder2009}.
Expanding the finite operator power and defining
$\tau_{ij}:=-(\im/2)(x_i-x_j)(p_i-p_j)$, the result takes the form
\begin{equation}
\begin{gathered}
\Psi_2(\bm x,\bm p)=\e^{\im\bm p\cdot\bm x}\,
R_N(\{\tau_{ij}\})\Bigl(\prod_{i<j}\tau_{ij}\Bigr)^{-1},\\[2pt]
R_2=1+\tau_{12},\qquad
R_3=\prod_{i<j}(1+\tau_{ij})+\tfrac12.
\end{gathered}
\label{eq:AIIalgebraic}
\end{equation}
For generic $N$, $R_N$ is a symmetric polynomial of degree at most one in each $\tau_{ij}$, with unit coefficient for $\prod_{i<j}\tau_{ij}$ [see $R_4$ in Eq.~\eqref{eq:em-aii-n4-R4} and $R_5, R_6$ in Ref.~\cite{SM}].
On the eigenvalue slice [Eq.~\eqref{eq:unified}], $\tau_{ij}=|z_i-z_j|^2/2$ and $\e^{\im\bm p\cdot\bm x}=\e^{-\sum_i|z_i|^2}$.
The denominator in Eq.~\eqref{eq:AIIalgebraic} cancels two powers of the Vandermonde in Eq.~\eqref{eq:unified}, giving

\begin{equation}
\rho^{\AIId}(\bm z)\propto
\e^{-\sum_i|z_i|^2}|\Delta(\bm z)|^{2}R_N.
\label{eq:main}
\end{equation}
Equation~\eqref{eq:main} is the Ginibre density dressed by $R_N$, which arises from the nonzero Calogero interaction and remains finite at eigenvalue collisions.
The pair-gas structure of class A does not survive the correction: the $+1/2$ in $R_3$ is an irreducible three-body interaction, and no pair potential reproduces $R_N$~\cite{SM}.
Since $R_N$ is finite at contact, the nearest-neighbor spacing distribution
$\pnn(s)$ [defined in Eq.~\eqref{eq:spacing}] behaves as $\pnn(s)\propto s^{3}$, not as the
$s^{5}$ suggested by the factor $|\Delta(\bm z)|^{\beta}$ of
Eq.~\eqref{eq:unified} at $\beta=4$ \cite{Grobe1989,Hamazaki2020,Akemann2022}.
Read off previously from $N=2$ surmises and from numerics, this $s^3$ law now follows from the exact density at every $N$.

By contrast, for class $\AId$ ($\beta=1$, $k=1/2$), $\mathcal H_k$ is not algebraically integrable, and we construct $\Psi_{1/2}$ as a noncompact orbital integral,
\begin{equation}
\Psi_{1/2}(\bm x,\bm p)
=[\Delta(\bm x)\Delta(-\im\bm p)]^{1/2}
\int_{\mathcal M_1}\dd\nu(G)\,
\e^{\im\Tr(PGXG^{-1})},
\label{eq:AIintegral}
\end{equation}
with $X=\operatorname{diag}(\bm x)$, $P=\operatorname{diag}(\bm p)$, and $\mathcal M_1$ the positive-Hermitian realization of $\mathrm{SO}(N,\C)/\mathrm{SO}(N)$.
The integral converges absolutely for
$\operatorname{Re}[-\im(p_i-p_j)(x_i-x_j)]>0$---in particular, on the
eigenvalue slice of Eq.~\eqref{eq:unified}---and real $\bm x,\bm p$ are
reached as boundary values.
Here, $\dd\nu(G)$ is the invariant measure on $\mathcal M_1$ induced by the holomorphic Haar form, with its overall normalization fixed by the free-wave condition $\Psi_{1/2}\sim\e^{\im\bm p\cdot\bm x}$ at large pair separations~\cite{SM}.
Earlier work on the $k=1/2$ system solved it by an orbital integral over the compact cycle \cite{Opdam1993,Brezin2003}, namely Eq.~\eqref{eq:AIintegral} with $\mathcal M_1$ replaced by $\mathrm O(N)$.
The two integrals solve the same differential equations with different boundary conditions, as the two kinds of Bessel functions do: the compact cycle selects the solution regular at the collisions $x_i=x_j$, while the noncompact cycle of Eq.~\eqref{eq:AIintegral} selects the free plane wave at infinity, the asymptotically free solution of Eq.~\eqref{eq:unified}~\cite{SM}.
Equation~\eqref{eq:unified} then determines $\rho^{\AId}(\bm z)$.

The hallmark of the class lies in its coupling.
At $k=1/2$, the coupling $2k(k-1)=-1/2$ of Eq.~\eqref{eq:HCal} reaches its minimum, making $\mathcal H_{1/2}$ the maximally attractive member of this Calogero family.
In the relative coordinate $u=x_i-x_j$, the pair potential is $-1/(4u^{2})$, precisely the critical strength of the inverse-square potential, beyond which $-\partial_u^{2}+g/u^{2}$ is unbounded below and the particles fall to the center~\cite{Case1950}.
For the eigenvalue problem with the pair potential $k(k-1)/u^2$, there exist two kinds of solutions (the modified Bessel functions of the first and second kinds), behaving as $u^{k}$ and $u^{1-k}$ at contact, respectively.
At criticality ($k=1/2$), the two exponents degenerate and logarithms appear (see End Matter for the two-body problem).
Correspondingly, the density acquires a logarithmic collision law, $\rho^{\AId}\sim|z_i-z_j|^{2}\log\bigl(1/|z_i-z_j|\bigr)$, giving $\pnn(s)\propto s^{3}\log(1/s)$, previously obtained from the $N=2$ surmise and from numerics \cite{Hamazaki2020,Akemann2022,Jaiswal2019} and established here at every $N$~\cite{SM}.

\emph{Differential equations for the density.---}
The strategy is to derive differential equations for the joint density $\rho\equiv\rho^{\mathcal C}$.
They follow from Gaussian integration by parts, treating all three classes at once.
Treat $H$ and $\bar H$ as independent variables, and let $(\nabla)_{ab}=\partial/\partial H_{ba}$ be the gradient within each class's matrix space; $\nabla$ inherits the symmetry of $H$: $\nabla=\nabla^{\mathrm T}$ for class $\AId$ and $\nabla=J\nabla^{\mathrm T}J^{-1}$ for class $\AIId$~\cite{SM}.
Associate with each trace power $q_\ell(H)=\Tr H^\ell$ the operator $D_{q_\ell}=\Tr(\nabla^\ell)$, up to a class normalization~\cite{SM}.
With the normalizations of Eq.~\eqref{eq:ensembles}, the Gaussian weight obeys
$\kappa\nabla\,\e^{-\Tr(H^\dagger H)/\kappa}=-\bar H^{\mathrm T}\,\e^{-\Tr(H^\dagger H)/\kappa}$
in every class, and $\bar H$ is inert under the holomorphic derivatives, so integrating by parts $\ell$ times gives, for every smooth symmetric function $F$ of the eigenvalues $\bm z(H)$,
\begin{equation}
\begin{split}
&\int\! D_{q_\ell}\bigl[F(\bm z(H))\bigr]\,\e^{-\Tr(H^\dagger H)/\kappa}\,\dd H\\
&\qquad=\int\! F(\bm z(H))\,q_\ell(\bar H)\,\e^{-\Tr(H^\dagger H)/\kappa}\,\dd H,
\end{split}
\label{eq:master}
\end{equation}
the $N$-body analogue of the single-variable Gaussian identity $\int\partial_zF\,\e^{-|z|^2}\dd^2z=\int\bar z\,F\,\e^{-|z|^2}\dd^2z$.

Both sides of Eq.~\eqref{eq:master} depend on $H$ only through its spectrum.
Write $H=SZS^{-1}$, with $Z=\operatorname{diag}(z_1,\ldots,z_N)$ and $S$ the diagonalizing similarity transformation.
On the right, $q_\ell(\bar H)=\sum_i\bar z_i^{\,\ell}=:q_\ell(\bar{\bm z})$.
On the left, $D_{q_\ell}$ commutes with conjugation by $S$ and can be evaluated at $Z$: $D_{q_\ell}[F(\bm z(H))]=(\rad{q_\ell}F)(\bm z)$, where $\rad{q_\ell}$ is the \emph{radial part}, a differential operator in the $z_i$ alone [Eq.~\eqref{eq:R2A}].
The matrix averages thus reduce to averages over the joint eigenvalue density $\rho$:
\begin{equation}
\int(\rad{q_\ell}F)(\bm z)\,\rho\;\dd\bm z
=\int F(\bm z)\,q_\ell(\bar{\bm z})\,\rho\;\dd\bm z
\label{eq:weak}
\end{equation}
for every smooth symmetric $F$, with $\dd\bm z:=\prod_i\dd^2z_i$.
Integrating by parts moves $\rad{q_\ell}$ onto $\rho$ (the definition of the transpose $\rad{q_\ell}^{T}$), and since $F$ is arbitrary the two integrands of Eq.~\eqref{eq:weak} agree pointwise:
\begin{equation}
\rad{q_\ell}^{T}(\bm z)\,\rho(\bm z,\bar{\bm z})
=q_\ell(\bar{\bm z})\,\rho(\bm z,\bar{\bm z}),
\qquad \ell=1,\dots,N .
\label{eq:strong}
\end{equation}
The transpose involves no complex conjugation, so Eq.~\eqref{eq:strong} is a joint eigenvalue problem in $\bm z$ at fixed $\bar{\bm z}$.
Since $\rho$ is real, complex conjugation supplies the equations with $\bm z$ and $\bar{\bm z}$ exchanged.

At the diagonal point $Z$, eigenvalue perturbation theory gives~\cite{SM}
\begin{equation}
\rad{q_1}=\sum_i\partial_{z_i},
\quad
\rad{q_2}=\sum_i\partial_{z_i}^2
+\beta\!\sum_{i<j}\frac{1}{z_i-z_j}
\bigl(\partial_{z_i}-\partial_{z_j}\bigr).
\label{eq:R2A}
\end{equation}
The pair term inherits the energy denominators of second-order perturbation
theory, and its coefficient $\beta=1,2,4$ counts the independent entries
coupling an eigenvalue pair in classes $\AId$, A, and $\AIId$.

Higher-order $\rad{q_\ell}$ can be calculated similarly, but ever more
laboriously.
Notably, the $\rad{q_\ell}$ inherit five properties from the
$D_{q_\ell}$---mutual commutativity, permutation symmetry, leading
derivatives $\sum_i\partial_{z_i}^{\ell}$, homogeneity of degree
$-\ell$, and rational coefficients with poles only at coincident
eigenvalues~\cite{SM}---which will identify them below without further
computation.

\emph{Solution by Calogero integrability.---}
Transposition gives $\rad{q_1}^{T}=-\rad{q_1}$; for $\rad{q_2}^{T}$, reordering derivatives and coefficients produces an inverse-square commutator term:

\begin{equation}
\rad{q_2}^{T}(\bm z)
=\sum_i\partial_{z_i}^2-\beta\!\sum_{i<j}\left[
\frac{\partial_{z_i}-\partial_{z_j}}{z_i-z_j}
-\frac{2}{(z_i-z_j)^2}\right].
\label{eq:R2AT}
\end{equation}
Next transform the density in Eq.~\eqref{eq:strong}, writing
$\rho=|\Delta|^{\beta}\psi$ with
$|\Delta|^{\beta}=\Delta^{k}\bar\Delta^{k}$, $k=\beta/2$, and $\bar\Delta$
inert under $\partial_{\bm z}$.
On the operators this is the gauge transformation
$\rad{q_\ell}^{T}\to\gauged{\ell}:=\Delta^{-k}\rad{q_\ell}^{T}\Delta^{k}$,
and Eq.~\eqref{eq:strong} becomes
$\gauged{\ell}\,\psi=q_\ell(\bar{\bm z})\,\psi$.

\begin{figure}[t]
\includegraphics[width=\columnwidth]{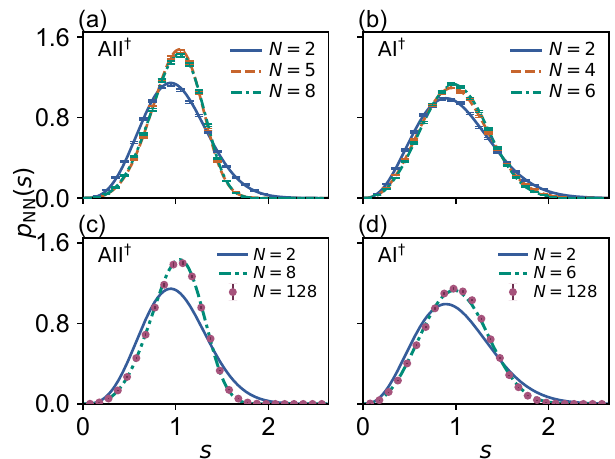}
\caption{\label{fig:spacing}(a),(b) Spacing distributions $\pnnN{N}(s)$, conditioned on an eigenvalue at the origin.
Lines: predictions from the exact densities.
Error bars: direct diagonalization.
(c),(d) Comparison with the bulk spacings of large random matrices; here the statistical errors are smaller than the markers.}
\end{figure}

Since $\gauged{1}=-\sum_i\partial_{z_i}$, $\psi$ carries total momentum $\sum_i\im\bar z_i$ [Eq.~\eqref{eq:unified}].
Moreover, $\gauged{2}=-\mathcal H_k|_{x_i\to z_i}$ identifies $\psi$ as an eigenfunction of the complexified Calogero Hamiltonian.
The higher $\gauged{\ell}$ need not be computed: transposition and
similarity preserve the five properties of $\rad{q_\ell}$, so each $\gauged{\ell}$
commutes with $\gauged{2}=-\mathcal H_k$ and retains the symmetry, homogeneity, and
rational coefficients, with leading derivatives
$(-1)^\ell\sum_i\partial_{z_i}^\ell$. These properties single out a unique
operator: $(-1)^\ell\gauged{\ell}=I_\ell$, the \emph{Liouville
integrals of motion} of the Calogero model $\mathcal H_k$
(see Supplemental Material~\cite{SM} for the explicit construction of
$I_\ell$).
Hence, $\psi$ is a joint eigenfunction of the complexified $I_\ell$, with the eigenvalues fixed by the momenta $\im\bar z_i$.
A boundary condition is needed to select the physical solution.
The eigenfunctions $\Psi_{\beta/2}(\bm z,\im\sigma\bar{\bm z})$, $\sigma\bar{\bm z}:=(\bar z_{\sigma(1)},\ldots,\bar z_{\sigma(N)})$ (particle
$z_i$ carrying the momentum $\im\bar z_{\sigma(i)}$ with $\sigma$ a permutation of the $N$ indices) solve the same equations.
The ensemble supplies the condition:
$\sum_i|z_i|^2\le\Tr(H^\dagger H)/\kappa$ forces $\rho$, and hence $\psi$,
to decay at least as fast as $\e^{-\sum_i|z_i|^2}$~\cite{SM}.
Every $\Psi_{\beta/2}(\bm z,\im\sigma\bar{\bm z})$ with
$\sigma\bar{\bm z}\neq\bar{\bm z}$ decays more slowly than this;
only $\Psi_{\beta/2}(\bm z,\im\bar{\bm z})$ survives, and
Eq.~\eqref{eq:unified} follows.

A Hermitian counterpart of Eq.~\eqref{eq:unified} is classical.
The Gaussian orthogonal, unitary, and symplectic ensembles have densities equal
to squared ground states of the \emph{trapped} Calogero model at the same
$k=\beta/2$ \cite{Calogero1971,Sutherland1971}, reached through
Dyson's Brownian motion~\cite{Dyson1962}, a dynamical mapping with
analogues in quantum transport~\cite{Beenakker1997,BeenakkerRejaei1993,Caselle1995}
and for non-Hermitian ensembles~\cite{Shukla2001}.
The relation found here differs from these.
The Hermitian density is the modulus squared of a single wave
function, the ground state; the non-Hermitian density \emph{is} a wave
function, and not one but a family: each configuration $\bm z$ is read in
its own scattering state, the one with momenta $\im\bar{\bm z}$.
Hermitian ensembles realize the bound states of the Calogero model;
non-Hermitian ensembles realize its scattering states.

\emph{Nearest-neighbor spacings.---}
We now use the exact joint eigenvalue distributions
[Eqs.~\eqref{eq:unified}, \eqref{eq:AIIalgebraic}, and \eqref{eq:AIintegral}]
to obtain finite-$N$ level-spacing statistics.
For each eigenvalue, define the unit-mean nearest-neighbor spacing
\begin{equation}
s_i=\min_{j\neq i}|z_i-z_j|/\bigl\langle\min_{j\neq i}|z_i-z_j|\bigr\rangle,
\label{eq:spacing}
\end{equation}
and let $\pnn(s)$ be its probability density.
At finite $N$, the mean density varies across the spectrum and no canonical
unfolding exists~\cite{Sa2020,Akemann2022}.
We avoid unfolding by conditioning on an eigenvalue at the origin and recording the nearest-neighbor distance $R_{\rm nn}=\min_{j\ge2}|z_j|$.
Integrating $\rho(0,z_2,\ldots,z_N)$ over those levels at fixed $R_{\rm nn}=s$,
and normalizing by the eigenvalue density at the origin, gives the
origin-conditioned density $P_N(s)$; rescaling to unit mean gives
$\pnnN{N}(s)=\langle R_{\rm nn}\rangle\,
P_N(\langle R_{\rm nn}\rangle\,s)$.
This differs from the global spacing at finite $N$, but is expected to
converge to it as $N\to\infty$, the origin being a bulk point of the
spectrum.

For class $\AIId$, the integrand is a polynomial times a Gaussian, so $P_N$ is
elementary at every $N$: the prefactor $s^3$ displays the cubic repulsion
[Eqs.~\eqref{eq:aii-n3-spacing} and \eqref{eq:em-aii-n4}; $5\le N\le8$ in
Ref.~\cite{SM}].
For class $\AId$, the same integral additionally contains the noncompact
orbital integral of Eq.~\eqref{eq:AIintegral} [Eqs.~\eqref{eq:em-ai-n2} and
\eqref{eq:em-ai-n3-palm}; $4\le N\le6$ in Ref.~\cite{SM}].

Figures~\ref{fig:spacing}(a,b) compare the exact origin-conditioned predictions with direct diagonalization.
Unlike the Hermitian Wigner surmise~\cite{Wigner1951,Mehta2004}, the non-Hermitian $N=2$ surmises are less accurate~\cite{Grobe1989,Hamazaki2020,Akemann2022,Jaiswal2019}.
Increasing to $N=8$ for class $\AIId$ and $N=6$ for class $\AId$ yields close agreement with the $N=128$ bulk distributions [Figs.~\ref{fig:spacing}(c,d)], providing useful finite-$N$ approximations to the bulk law.

\emph{Deviation from Coulomb gas.---}
We further study the long-distance 
behavior of the spectral correlations in the
bulk, described by the two-point correlation function
\begin{equation}
 \rho_2(z_1,z_2)=\Bigl\langle\sum_{i\neq j}
 \delta(z_1-z_i)\,\delta(z_2-z_j)\Bigr\rangle
 \label{eq:pair-correlation-def}
\end{equation}
and its connected part $\rho_2^{(c)}(r)=\rho_2/\rho_1^{2}-1$, with $\rho_1$ the mean density and $r=|z_1-z_2|$.
For the Ginibre ensemble (class A), it is known that
$\rho_{2,\mathrm A}^{(c)}=-\e^{-r^{2}}$ at
$\rho_1=1/\pi$~\cite{Ginibre1965,ByunForrester2024}. The Gaussian decay originates from the Coulomb-gas analogue.
The joint density is a Boltzmann weight, $\rho^{\mathrm A}\propto\e^{-\beta U}$
with $U=\beta^{-1}\sum_i|z_i|^{2}-\sum_{i<j}\ln|z_i-z_j|$ and $\beta=2$:
the energy of two-dimensional charges, repelling through the Coulomb
potential $-\ln r$ and confined by a uniform neutralizing background.
Such a Coulomb gas is a conductor and features perfect screening: every
charge is neutralized by its screening cloud, and correlations are extremely short
ranged~\cite{StillingerLovett1968,Martin1988}.

\begin{figure}[t]
\includegraphics[width=0.8\columnwidth]{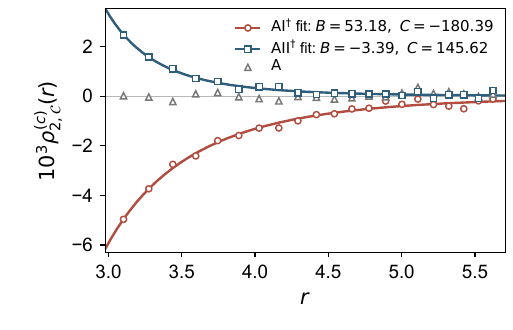}
\caption{\label{fig:long-distance}Bulk pair correlation
$\rho_{2,\mathcal C}^{(c)}(r)$ at separation $r$, after rescaling each class to $\rho_1=1/\pi$, for $N=128$,
from $3\times10^6$ matrices per class.  Points use the pair-midpoint
event-mixing estimator in Eq.~\eqref{eq:em-pair-estimator}; solid curves fit
$A r^{-6}+B r^{-8}+C r^{-10}$, with $A=-8$ ($\AId$) or $1$ ($\AIId$)
fixed and $B,C$ fitted.  Error bars are one standard error
over independent matrix blocks.}
\end{figure}

Perfect screening fails in classes $\AId$ and $\AIId$.
To investigate the long-range behavior of the joint densities, we apply the
WKB approximation to the Calogero Hamiltonian $\mathcal H_k$~\cite{LandauLifshitz1977,Polychronakos2006}.
For well-separated particles with momenta $\bm p$, the WKB phase $\sigma$
accumulated at position $\bm x$ is given by
\begin{equation}
 \sigma=\sum_{i<j}\int_{u_{ij}}^{\infty}\!\dd u
 \biggl[\,q_{ij}-\sqrt{q_{ij}^{2}-\frac{k(k-1)}{u^{2}}}\,\biggr]
 \simeq\sum_{i<j}\frac{k(k-1)}{2\,q_{ij}u_{ij}},
 \label{eq:wkb-phase}
\end{equation}
with $u_{ij}=x_i-x_j$, $q_{ij}=(p_i-p_j)/2$, and $k(k-1)/u^{2}$ the pair
potential in the relative coordinate.
Inserting $\Psi_k=\e^{\im(\bm p\cdot\bm x+\sigma)}$ into
Eq.~\eqref{eq:unified}, the density takes the asymptotic form
$\rho^{\mathcal C}\propto\e^{-\beta U}$ with
\begin{equation}
 U=\frac{\sum_i|z_i|^{2}}{\beta}
 -\sum_{i<j}\Bigl[\ln|z_i-z_j|+\frac{k-1}{2|z_i-z_j|^{2}}\Bigr]+\cdots.
 \label{eq:long-distance-jost}
\end{equation}
The Fourier transform of $\rho_2^{(c)}$ defines the structure factor
$S(q)\equiv1+\rho_1\!\int\!\dd^2r\,\e^{-\im\bm q\cdot\bm r}\rho_2^{(c)}(r)$.
It can be evaluated by the random-phase approximation,
$S(q)\approx[1+\beta\rho_1\hat v(q)]^{-1}$, where $\hat v(q)$ is the Fourier
transform of the two-body interaction~\cite{HansenMcDonald2013}.
For the Coulomb potential, $\hat v(q)=2\pi/q^{2}$; the correction
$-(k-1)/(2r^{2})$ additionally gives
$\delta\hat v(q)=-\pi(k-1)\ln(q_0/q)$, which diverges logarithmically in
the infrared, with $q_0$ the corresponding cutoff.
Expanding at small $q$,
\begin{equation}
 S(q)=\frac{q^{2}}{4}+\frac{k-1}{8}\,q^{4}\ln\frac{q_0}{q}+\cdots.
 \label{eq:screened-tail}
\end{equation}
Transforming the definition of $S(q)$ back to real space, the analytic terms
decay rapidly, while the singular one, $q^{4}\ln(q_0/q)$, transforms to
$32/(\pi r^{6})$.
With $\rho_1=1/(\pi k)$ in the normalization of Eq.~\eqref{eq:ensembles},
this gives $\rho_2^{(c)}(r)\simeq4k(k-1)/r^{6}$; rescaling $r$ by
$\sqrt{k}$ brings every class to the bulk density $1/\pi$, where
\begin{equation}
 \rho_{2,\AId}^{(c)}(r)\simeq-\frac{8}{r^{6}},
 \qquad
 \rho_{2,\AIId}^{(c)}(r)\simeq\frac{1}{r^{6}},
 \label{eq:long-distance-tail}
\end{equation}
power-law tails whose sign and amplitude are fixed by the symmetry
through $k(k-1)$: anticorrelation for $\AId$ ($k<1$) and correlation for
$\AIId$ ($k>1$).
Diagonalization at $N=128$ is consistent with these tails after including
subleading finite-distance corrections
[Fig.~\ref{fig:long-distance}, End Matter].

\emph{Summary and outlook.---}
We have obtained the exact joint eigenvalue densities of the Gaussian non-Hermitian ensembles in classes $\AId$ and $\AIId$, at arbitrary matrix size $N$ [Eqs.~\eqref{eq:unified}, \eqref{eq:main}, and \eqref{eq:AIintegral}].
We derive differential equations for the joint density and identify their Calogero integrable structure: the density is a scattering state.
The model assigns the three classes three levels of structure: free for class A, algebraically integrable for class $\AIId$, and Liouville integrable for class $\AId$.
To our knowledge, the previously known joint densities in the Gaussian ensembles are pair gases, with eigenvalues interacting through two-body potentials alone~\cite{Mehta2004,Osborn2004,KawabataRyu2026,Ginibre1965,ByunForrester2024,Lehmann1991,ForresterNagao2007,Forrester2010,Akemann2005,AkemannPhillipsSommers2010,Kieburg2026}; the densities obtained here are the first with irreducible many-eigenvalue interactions, a step in understanding exotic spectral correlations in RMT.
We have verified the results numerically, through the level spacings of finite-$N$ random matrices and the long-range power law decay of two-point spectral correlation function.
We have found both classes realized in the reflection matrix of a Hermitian disordered conductor, whose transposition symmetry follows from the time-reversal symmetry of the sample (End Matter).

The exact densities provide the starting point for deriving bulk and edge correlation functions, with controlled large-$N$ limits.
Since the Calogero equations exist at every real coupling $k$, they also
suggest an analogue of the Hermitian $\beta$ ensembles~\cite{Forrester2010,DumitriuEdelman2002}
(i.e., allowing generic positive $\beta$) for non-Hermitian random matrices.
Our joint densities may also find applications in quantum Hall physics~\cite{ByunForrester2024,Zabrodin2003,Bourgine2024,Borutta2026}.

{\it Acknowledgments.---}
Z.X. thanks Kohei Kawabata for helpful discussions.
The computations reported in this paper were performed using Princeton Research Computing resources.
Z.X. is supported by the Princeton Quantum Initiative Fellowship.
S.R. is supported by a Simons
Investigator Grant from the Simons Foundation 
(Award No.\ 566116). This work is supported by the Gordon and Betty
Moore Foundation EPiQS initiative, Grant GBMF8685.01.
The authors used OpenAI Codex (GPT-5.6) and Claude Code (Opus 5) for code development and debugging, checks of analytical steps, and manuscript revision.
The authors critically reviewed and independently verified all outputs through analytical and numerical checks, and retain full responsibility for this work.

\textit{Note added.}---A companion work, appearing concurrently on arXiv, uses replica nonlinear $\sigma$ models to study level correlations~\cite{Chen2026}.
In a forthcoming work, we generalize the formalism to more non-Hermitian symmetry classes in the 38-fold classification.

\bibliography{references}

\onecolumngrid
\vspace{6pt}
\begin{center}
\rule{0.45\textwidth}{0.4pt}\\[5pt]
\textbf{End Matter}
\end{center}
\twocolumngrid
\newcommand{\letterappendix}[2]{
  \setcounter{equation}{0}
  \renewcommand{\theequation}{#1\arabic{equation}}
  \renewcommand{\theHequation}{#1.\arabic{equation}}
  \section*{Appendix #1: #2}
}

\letterappendix{A}{The two-body problem}
At $N=2$, the Calogero model is solvable at every coupling.
Separating the centre of mass, with $u=x_1-x_2$, $q=(p_1-p_2)/2$, and
$\Psi_k=\e^{\im(p_1+p_2)(x_1+x_2)/2}\psi_k(u,q)$, the relative part of
Eq.~\eqref{eq:HCal} is $-2\partial_u^{2}+2k(k-1)/u^{2}$ with eigenvalue
$2q^{2}$; dividing by $2$,
\begin{equation}
h_k\,\psi_k=q^{2}\psi_k ,
\qquad
h_k=-\partial_u^{2}+\frac{k(k-1)}{u^{2}} .
\label{eq:em-hrel}
\end{equation}
With $\nu=k-1/2$, so that $\nu^{2}-1/4=k(k-1)$, the substitution
$\psi=\sqrt{u}\,Z(qu)$ turns Eq.~\eqref{eq:em-hrel} into Bessel's equation,
\begin{equation}
w^{2}Z''+wZ'+\bigl(w^{2}-\nu^{2}\bigr)Z=0 ,\qquad w=qu .
\label{eq:em-bessel}
\end{equation}
Its two solutions give
\begin{equation}
\psi^{\rm reg}(u)=\sqrt{u}\,J_{\nu}(qu),
\qquad
\psi^{\rm out}(u)=\sqrt{u}\,H^{(1)}_{\nu}(qu),
\label{eq:em-twosol}
\end{equation}
with $J_\nu$ the Bessel function of the first kind and $H^{(1)}_\nu$ the
Hankel function~\cite{Olver2010,AbramowitzStegun1964}.
At the collision, $\psi^{\rm reg}\sim u^{k}$ and $\psi^{\rm out}\sim u^{1-k}$;
at large $qu$, $\psi^{\rm reg}$ is a standing wave and $\psi^{\rm out}$ is
purely outgoing.
These are the two boundary conditions of the main text.
At the imaginary momenta of Eq.~\eqref{eq:unified}, $q=\im\bar q$, they
become the modified Bessel functions, $J_\nu\to I_\nu$ growing and
$H^{(1)}_\nu\to K_\nu$ decaying~\cite{Olver2010}.
Normalized to a unit outgoing wave,
\begin{equation}
\psi_k(u,q)=\im^{k}\sqrt{\tfrac{\pi qu}{2}}\;H^{(1)}_{k-\frac12}(qu)
\;\xrightarrow[\;qu\to\infty\;]{}\;\e^{\im qu} .
\label{eq:em-hankel}
\end{equation}

The three classes have $\nu=1/2$, $3/2$, and $0$.
At half-integer $\nu$, the Bessel functions are elementary:
$\psi_1=\e^{\im qu}$ at $k=1$, and $\psi_2=\e^{\im qu}(1+\im/qu)$ at $k=2$,
which is Eq.~\eqref{eq:AIIalgebraic} at $N=2$ with $\tau_{12}=-\im qu$ and
$R_2=1+\tau_{12}$.
At $\nu=0$ (class $\AId$), the two exponents $u^{k}$ and $u^{1-k}$ coincide
at $\sqrt{u}$; a second solution with the same power cannot exist, and a
logarithm appears instead, $\psi^{\rm out}\sim\sqrt{u}\log u$.
This is the origin of the logarithmic collision law of the main text.

\letterappendix{B}{The Baker--Akhiezer formula and explicit spacing distributions}
The class-$\AIId$ scattering state is given in closed form at every
$N$ by the operator formula~\cite{Chalykh1999,Felder2009}
\begin{equation}
\Psi_2(\bm x,\bm p)=
\frac{\bigl(\mathcal L+{\textstyle\sum_i}p_i^{2}\bigr)^{\!N(N-1)/2}
\bigl[\Delta(\bm x)^{2}\,\e^{\im\bm p\cdot\bm x}\bigr]}
{2^{N(N-1)/2}\,\bigl[N(N-1)/2\bigr]!\,\Delta(\bm x)\,\Delta(\im\bm p)},
\label{eq:AIIberest}
\end{equation}
where $\mathcal L=\sum_i\partial_{x_i}^{2}-2\sum_{i<j}(x_i-x_j)^{-1}(\partial_{x_i}-\partial_{x_j}){}=-\Delta(\bm x)\,\mathcal H_2\,\Delta(\bm x)^{-1}$, with $\mathcal H_2$ the Hamiltonian of Eq.~\eqref{eq:HCal} at $k=2$.
Expanding the finite operator power yields Eq.~\eqref{eq:AIIalgebraic} of the main text.

All spacing formulas below are in the eigenvalue normalization of
Eq.~\eqref{eq:ensembles}, before the rescaling to unit mean.

For class $\AIId$ at $N=3$, the origin-conditioned spacing density is
\begin{equation}
P_{3}^{\AIId}(s)=\frac{s^{3}\e^{-2s^{2}}}{108}
\bigl(3s^{10}\!+23s^{8}\!+74s^{6}\!+136s^{4}\!+138s^{2}\!+60\bigr).
\label{eq:aii-n3-spacing}
\end{equation}
At $N=4$, the polynomial of
Eq.~\eqref{eq:AIIalgebraic} is
\begin{equation}
R_4=\prod_{i<j}(1+\tau_{ij})
+\frac12\sum_{v=1}^{4}\prod_{j\neq v}(1+\tau_{vj})
+\frac14\sum_{i<j}(1+\tau_{ij}),
\label{eq:em-aii-n4-R4}
\end{equation}
with all indices running over $1,\ldots,4$; on the eigenvalue slice,
$\tau_{ij}=|z_i-z_j|^2/2$, and together with Eq.~\eqref{eq:main}
this is the exact $N=4$ joint density.
The corresponding origin-conditioned spacing density is
\begin{multline}
P_{4}^{\AIId}(s)=\frac{s^{3}\e^{-3s^{2}}}{1036800}\bigl(483840+1486080s^{2}\\
{}+2273280s^{4}+2275200s^{6}+1646496s^{8}\\
{}+902512s^{10}+383520s^{12}+127116s^{14}\\
{}+32412s^{16}+6076s^{18}+750s^{20}+45s^{22}\bigr).
\label{eq:em-aii-n4}
\end{multline}
The explicit $5\le N\le8$ densities are given in the Supplemental Material~\cite{SM}.

For class $\AId$ at $N=2$,
\begin{equation}
P_{2}^{\AId}(s)=\tfrac{3}{2}\,s^{3}\,
\e^{-s^{2}/2}K_{0}\!\bigl(s^{2}/2\bigr),
\label{eq:em-ai-n2}
\end{equation}
with the critical logarithm $s^{3}\log(1/s)$ at short distance.
At $N=3$, the exact joint density is, with $a,b,c$ the three squared
pair distances and $K(m)$ the complete elliptic integral of the first
kind,
\begin{equation}
\rho_{3}^{\AId}(\bm z)=\frac{2abc}{3\pi^{7/2}}\,
\e^{-\sum_{i}|z_{i}|^{2}}\,\mathcal E(a,b,c),
\label{eq:em-ai-n3-jpdf}
\end{equation}
\begin{equation}
\mathcal E(a,b,c)=\int_{0}^{\infty}\!\dd p\,
\frac{\e^{-p}}{\sqrt{Q(p)}}\,
K\!\left(1-\frac{abc}{Q(p)}\right),
\label{eq:em-elliptic}
\end{equation}
with $Q(p)=(p+a)(p+b)(p+c)$.
The origin-conditioned spacing density follows by fixing the reference
eigenvalue at the origin and the remaining two at moduli $s\le t$ with
relative angle $\theta$, so that $a=s^{2}$, $b=t^{2}$,
$c=a+b-2st\cos\theta$:
\begin{equation}
P_{3}^{\AId}(s)=\frac{16s}{3\pi^{3/2}}
\int_{s}^{\infty}\!t\,\dd t\int_{0}^{2\pi}\!\dd\theta\;
abc\,\e^{-a-b}\,\mathcal E(a,b,c).
\label{eq:em-ai-n3-palm}
\end{equation}

\letterappendix{C}{Numerical test of the long-distance tails}
We diagonalize $3\times10^{6}$ independent Gaussian matrices from
Eq.~\eqref{eq:ensembles} at $N=128$ in each class, retaining one eigenvalue
per Kramers doublet in class $\AIId$, and rescale the spectra to
$\rho_1=1/\pi$.  For each pair we record its separation
$r=|z_i-z_j|$ and midpoint $m=(z_i+z_j)/2$, retaining only midpoints in the
central bulk.  We group the separations into common radial bins.
The counts $n_{{\rm same},\ell}$ and $n_{{\rm cross},\ell}$ in bin $\ell$
come, respectively, from pairs within one spectrum and from pairs drawn from
independent spectra.  Their individual separations need not coincide.
After the corresponding combinatorial normalization, the binwise estimator is
\begin{equation}
 1+\widehat\rho_{2,\ell}^{(c)}
 =\frac{n_{{\rm same},\ell}}{n_{{\rm cross},\ell}}.
 \label{eq:em-pair-estimator}
\end{equation}
The cross-matrix histogram is the uncorrelated bulk reference and removes
the annular phase-space factor common to the two histograms.
We estimate errors from independent matrix blocks, keeping all pairs from a
given matrix in the same block.  In Fig.~\ref{fig:long-distance}, we fit the
$21$ radial-bin estimates to $A r^{-6}+B r^{-8}+C r^{-10}$ using their full
covariance matrix.  The leading coefficient is fixed to $A=-8$ for class
$\AId$ and $A=1$ for class $\AIId$, as predicted by
Eq.~\eqref{eq:long-distance-tail}; only $B$ and $C$ are fitted.
The fit is evaluated at the bin centers, $3.10\le r\le5.62$.
The fitted $(B,C)$ are $(53.18,-180.39)$ and $(-3.39,145.62)$,
with $\chi^2=18.27$ and $8.31$ for $19$ degrees of freedom, respectively.

\begin{figure}[t]
\includegraphics[width=\columnwidth]{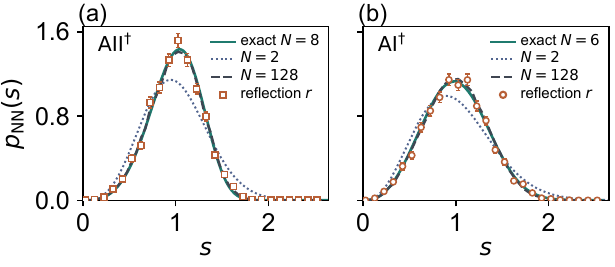}
\caption{\label{fig:reflection}Spacing distribution of the complex eigenvalues of the reflection block $r$, locally unfolded to unit mean, for (a) class $\AIId$ and (b) class $\AId$.
Error bars: standard errors over $400$ disorder realizations.}
\end{figure}

\letterappendix{D}{Emergence in Hermitian quantum transport}
The two TRS$^{\dagger}$ classes arise in transport through entirely
Hermitian systems.  We consider disordered cubic lattices of $L^3$ sites,
\begin{equation}
\begin{aligned}
H_{\mathrm{AI}}&=\sum_i v_i\,c_i^\dagger c_i
 -\sum_{i,\mu}\left(c_i^\dagger c_{i+\hat\mu}+\mathrm{H.c.}\right),\\
H_{\mathrm{AII}}&=\sum_i v_i\,c_i^\dagger\sigma_0 c_i
 -\sum_{i,\mu}\left[c_i^\dagger\bigl(\sigma_0+\im\lambda\sigma_\mu\bigr)c_{i+\hat\mu}
 +\mathrm{H.c.}\right],
\end{aligned}
\label{eq:scattering-hamiltonian}
\end{equation}
where $c_i$ annihilates a spinless (spin-$1/2$) fermion at site $i$,
$\lambda$ is the spin-orbit coupling, the onsite potentials $v_i$ are
independent and uniform on $[-W/2,W/2]$, and the hopping is set to unity.
Both are time-reversal symmetric, $\mathcal T_+H^{*}\mathcal T_+^{-1}=H$,
with $\mathcal T_+=1$ and $\mathcal T_+=\im\sigma_y$: the Wigner--Dyson
classes AI and AII.
Ideal leads are attached to the two $x$ surfaces, and $r$ denotes the
left-reflection block.
For a two-terminal geometry, the scattering matrix at energy $E$
follows from the Mahaux--Weidenm\"uller formula \cite{Beenakker1997},
\begin{equation}
 S(E)=\frac{1-\im\pi K(E)}{1+\im\pi K(E)},
 \qquad
 K(E)=\mathcal W^{\dagger}\frac{1}{E-H}\,\mathcal W,
 \label{eq:MW}
\end{equation}
where $\mathcal W$ couples the sample to the leads.
Writing $S=\left(\begin{smallmatrix}r&t'\\t&r'\end{smallmatrix}\right)$,
the block $r$ ($r'$) reflects the left (right) lead onto itself.
If $H$ has TRS, $H=\mathcal T_+H^{*}\mathcal T_+^{-1}$, and the leads
respect it, $\mathcal T_+\mathcal W^{*}\mathcal T_+^{-1}\propto\mathcal W$,
then $\mathcal T_+K^{\mathrm T}\mathcal T_+^{-1}=K$ and hence
$\mathcal T_+S^{\mathrm T}\mathcal T_+^{-1}=S$.
Because $\mathcal T_+$ acts locally in space, the constraint descends to
the reflection block alone,
\begin{equation}
 \mathcal C_+\,r^{\mathrm T}\mathcal C_+^{-1}=r,
 \label{eq:refl-symmetry}
\end{equation}
with $\mathcal C_+$ the restriction of $\mathcal T_+$ to the channels of one lead.
This is TRS$^{\dagger}$ [Eq.~\eqref{eq:trs}], and it inherits the sign:
classes AI and AII produce $\AId$ and $\AIId$ for
$r$~\cite{Kawabata2023SVD}.

We take $L=12$ with periodic transverse boundaries, $E=0.1$, and $W=1.6$
(class AI) or $(W,\lambda)=(0.8,0.2)$ (class AII).
Microscopically, $r$ is not a Gaussian matrix: unitarity of $S$ confines
its spectrum to the unit disk.
The reflection-matrix spacings computed with Kwant~\cite{Groth2014}
nevertheless follow the $N=8$ class-$\AIId$ and $N=6$ class-$\AId$
predictions (Fig.~\ref{fig:reflection}); numerical details are given in
Ref.~\cite{SM}.

\end{document}


\title{Supplemental Material for\texorpdfstring{\\}{: } ``Exact joint eigenvalue densities of non-Hermitian random matrices are Calogero scattering states''}

\author{Zhenyu Xiao\equalcontrib}
\email{zyxiao@princeton.edu}
\affiliation{Princeton Quantum Initiative, Princeton University, Princeton, New Jersey 08544, USA}
\affiliation{Princeton Center for Theoretical Science, Princeton University, Princeton, New Jersey 08544, USA}
\author{Ze Chen\equalcontrib}
\affiliation{Department of Physics, Princeton University, Princeton, New Jersey 08544, USA}
\affiliation{Department of Electrical and Computer Engineering, Princeton University, Princeton, New Jersey 08544, USA}
\author{Shinsei Ryu}
\email{shinseir@princeton.edu}
\affiliation{Department of Physics, Princeton University, Princeton, New Jersey 08544, USA}

\maketitle
\makeatletter
\renewcommand*\l@section{\@dottedtocline{1}{0pt}{3.0em}}
\renewcommand*\l@subsection{\@dottedtocline{2}{1.6em}{2.4em}}
\makeatother
\tableofcontents

This Supplemental Material contains the proofs underlying the Letter.
Section~\ref{sec:calogero-model} introduces the rational Calogero model,
its integrals of motion, and their uniqueness.
Section~\ref{sec:spectral-system} derives the differential equations for
the joint density and solves them.
Section~\ref{sec:polar} derives the densities of classes A and $\AId$
directly from the matrix measure, and gives the orbit interpretation for
class $\AIId$.
Sections~\ref{sec:aii-solution} and \ref{sec:ai-scattering} construct the
solutions of classes $\AIId$ and $\AId$ and evaluate their spacing
distributions.
Section~\ref{sec:spacing} describes the numerical checks, and
Section~\ref{sec:transport} the reflection-matrix simulations.

\section{The rational Calogero model and its integrals of motion}
\label{sec:calogero-model}

This section assembles the Calogero-model input used in the proof: the
model itself, the Dunkl-operator construction of its integrals of motion,
and a uniqueness characterization of those integrals.
All operators act on functions of the complex variables
$\bm z=(z_1,\ldots,z_N)$, obtained from the real Calogero coordinates of
the main text by the complexification $x_i\to z_i$.

\subsection{The model}
The rational Calogero Hamiltonian at coupling $k$ is
\begin{equation}
 \mathcal H_k=-\sum_i\partial_{z_i}^2
 +\sum_{i<j}\frac{2k(k-1)}{(z_i-z_j)^2}.
 \label{eq:calogero-def}
\end{equation}
The interaction vanishes at $k=0$ and $k=1$; for every real $k$, the
model is Liouville integrable: it possesses $N$ commuting integrals of
motion $I_\ell$, one for each power of the momenta, with $-I_2=\mathcal H_k$
\cite{Olshanetsky1983}.
At integer $k$, the model is in addition \emph{algebraically
integrable}, and its asymptotically free eigenfunction is a
Baker--Akhiezer function: a plane wave times a rational function
\cite{Chalykh1990,Chalykh1999,Felder2009}.

\subsection{Dunkl operators and the integrals of motion}
The integrals of motion are constructed from the Dunkl operators
\begin{equation}
 T_i=\partial_{z_i}
 +k\sum_{j\neq i}\frac{1-s_{ij}}{z_i-z_j},
 \qquad i=1,\ldots,N,
 \label{eq:dunkl-sm}
\end{equation}
where $s_{ij}$ exchanges the two arguments $z_i$ and $z_j$,
\begin{equation}
 (s_{ij}f)(z_1,\ldots,z_i,\ldots,z_j,\ldots,z_N)
 =f(z_1,\ldots,z_j,\ldots,z_i,\ldots,z_N).
 \label{eq:sij-def}
\end{equation}
The Dunkl operators commute pairwise,
$[T_i,T_j]=0$~\cite{Dunkl1989,Polychronakos1992,Polychronakos2006,Correa2014}.
The power sums $\sum_iT_i^\ell$ have three properties that we use:
\begin{enumerate}
\item[(a)] they are permutation invariant, and therefore preserve the
space of symmetric functions;
\item[(b)] on symmetric functions, each acts as an ordinary differential
operator: normal-ordering the reflections to the right and evaluating
them there ($s_{ij}\to1$ on symmetric functions) removes all reflections;
we denote the resulting operator by $\mathrm{res}(\sum_iT_i^\ell)$;
\item[(c)] the $\mathrm{res}(\sum_iT_i^\ell)$ commute among themselves,
because the $T_i$ do.
\end{enumerate}
The two lowest members follow directly from Eq.~\eqref{eq:dunkl-sm}.
At first order, the reflection terms annihilate symmetric functions,
$(1-s_{ij})F=0$, so $\mathrm{res}(\sum_iT_i)=\sum_i\partial_{z_i}$.
At second order, act with $T_i$ twice on a symmetric $F$.
The first action again gives $T_iF=\partial_{z_i}F$; but the intermediate
result $\partial_{z_i}F$ is no longer symmetric, and the reflections now
act nontrivially: swapping $z_i$ and $z_j$ in $\partial_{z_i}F$ gives
$s_{ij}(\partial_{z_i}F)=\partial_{z_j}F$ (because $F$ itself is
symmetric), so $(1-s_{ij})\partial_{z_i}F
=(\partial_{z_i}-\partial_{z_j})F$ and
\begin{equation}
 T_i(T_iF)
 =\partial_{z_i}^2F
 +k\sum_{j\neq i}
 \frac{(\partial_{z_i}-\partial_{z_j})F}{z_i-z_j}.
 \label{eq:dunkl-second-action}
\end{equation}
Summing over $i$, the $(i,j)$ and $(j,i)$ terms of the interaction
coincide, and
\begin{equation}
 \mathrm{res}\!\left(\sum_iT_i^2\right)
 =\sum_i\partial_{z_i}^2
 +2k\sum_{i<j}\frac{1}{z_i-z_j}
 \bigl(\partial_{z_i}-\partial_{z_j}\bigr).
 \label{eq:dunkl-quadratic}
\end{equation}
As Eq.~\eqref{eq:dunkl-second-action} shows, the ordering is derivative
first: the derivatives act on the function, and the result is divided by
$z_i-z_j$; the same convention applies throughout.
Now define the gauged family, with
$\Dl(\bm z)=\prod_{i<j}(z_i-z_j)$ the Vandermonde determinant,
\begin{equation}
 I_\ell:=\Dl^{k}\,\mathrm{res}\!\left(\sum_iT_i^\ell\right)\Dl^{-k},
 \qquad \ell=1,2,\ldots.
 \label{eq:integrals-def}
\end{equation}
For $\ell=1$, $\sum_i\partial_{z_i}\log\Dl
=\sum_i\sum_{j\neq i}(z_i-z_j)^{-1}=0$ by antisymmetry, so
\begin{equation}
 I_1=\sum_i\partial_{z_i},
 \label{eq:integral-one}
\end{equation}
$\im$ times the total momentum.
For $\ell=2$, the conjugation is evaluated with the exact Vandermonde
identity
\begin{equation}
 \Dl^{k}\Bigl[\sum_i\partial_{z_i}^2
 +2k\sum_{i<j}\frac{1}{z_i-z_j}
 \bigl(\partial_{z_i}-\partial_{z_j}\bigr)\Bigr]\Dl^{-k}
 =\sum_i\partial_{z_i}^2
 -2k(k-1)\sum_{i<j}\frac1{(z_i-z_j)^2},
 \label{eq:calogero-gauge}
\end{equation}
checked by expanding both sides, and gives
\begin{equation}
 -I_2
 =-\sum_i\partial_{z_i}^2
 +2k(k-1)\sum_{i<j}\frac1{(z_i-z_j)^2}
 =\mathcal H_k .
 \label{eq:calogero-hamiltonian}
\end{equation}
The $I_\ell$ therefore form a commuting family of differential operators
that contains the Calogero Hamiltonian: they are the quantum Liouville
integrals of motion of the rational Calogero model
\cite{Olshanetsky1983}, and Eq.~\eqref{eq:integrals-def} is their standard
construction.

\subsection{Uniqueness of the integrals of motion}
\label{sec:rigidity}
In this subsection, we show that the integrals of motion are rigid.
Call a \emph{reference operator} any operator of the form
$\sum_i\partial_{z_i}^2$ plus lower-order terms with rational
coefficients whose poles lie on the collision hyperplanes; both
$\mathrm{res}(\sum_iT_i^2)$ and the complexified Hamiltonian $-\mathcal H_k=I_2$
qualify.
If two differential operators $L$ and $L'$ commute with a common
reference operator, have the same principal part
$\sum_i\partial_{z_i}^{\ell}$, are homogeneous of degree $-\ell$, and
have rational coefficients with poles only on the collision hyperplanes,
then $L=L'$.

Consider $B=L-L'$.
Because the principal parts cancel, $B$ is a differential operator of some
lower order $r<\ell$: its terms with exactly $r$ derivatives read
\begin{equation}
 B_{\mathrm{top}}
 =\sum_{\substack{r_1,\ldots,r_N\ge0\\ r_1+\cdots+r_N=r}}
 b_{r_1\cdots r_N}(\bm z)\,
 \partial_{z_1}^{r_1}\cdots\partial_{z_N}^{r_N},
 \label{eq:rigidity-top}
\end{equation}
with one coefficient function for each way of distributing the $r$
derivatives among the $N$ variables.
Now evaluate the commutator $[\mathrm{res}(\sum_iT_i^2),B]=0$ on the
trial wave $\e^{\lambda\bm w\cdot\bm z}$,
$\bm w\cdot\bm z=\sum_iw_iz_i$, and expand for $\lambda\to\infty$.
Each derivative acting on the exponential alone produces its factor
$\lambda w_i$ exactly, so
\begin{equation}
 B\,\e^{\lambda\bm w\cdot\bm z}
 =\e^{\lambda\bm w\cdot\bm z}\sum_{m\le r}\lambda^{m}\,b_m(\bm z,\bm w),
 \qquad
 b_m(\bm z,\bm w)
 :=\sum_{r_1+\cdots+r_N=m}b_{r_1\cdots r_N}(\bm z)\,
 w_1^{r_1}\cdots w_N^{r_N},
 \label{eq:rigidity-symbol}
\end{equation}
and likewise, by Eq.~\eqref{eq:dunkl-quadratic},
\begin{equation}
 \mathrm{res}\Bigl(\sum_iT_i^2\Bigr)\e^{\lambda\bm w\cdot\bm z}
 =\e^{\lambda\bm w\cdot\bm z}
 \Bigl[\lambda^2\sum_iw_i^2
 +2k\lambda\sum_{i<j}\frac{w_i-w_j}{z_i-z_j}\Bigr].
 \label{eq:rigidity-wave}
\end{equation}
Apply the two orderings to the trial wave and use the product rule,
with $f:=\sum_m\lambda^mb_m$:
\begin{align}
 \mathrm{res}\Bigl(\sum_iT_i^2\Bigr)
 \bigl(B\,\e^{\lambda\bm w\cdot\bm z}\bigr)
 &=\e^{\lambda\bm w\cdot\bm z}
 \Bigl[\lambda^2\Bigl(\sum_iw_i^2\Bigr)f
 +2\lambda\sum_iw_i\,\partial_{z_i}f
 +2k\lambda\sum_{i<j}\frac{w_i-w_j}{z_i-z_j}\,f\notag\\
 &\hspace{7em}
 +\sum_i\partial_{z_i}^2f
 +2k\sum_{i<j}\frac{1}{z_i-z_j}
 \bigl(\partial_{z_i}-\partial_{z_j}\bigr)f\Bigr],
 \label{eq:rigidity-order1}\\
 B\Bigl(\mathrm{res}\Bigl(\sum_iT_i^2\Bigr)
 \e^{\lambda\bm w\cdot\bm z}\Bigr)
 &=\e^{\lambda\bm w\cdot\bm z}
 \Bigl[\Bigl(\lambda^2\sum_iw_i^2
 +2k\lambda\sum_{i<j}\frac{w_i-w_j}{z_i-z_j}\Bigr)f
 +O(\lambda^{r})\Bigr],
 \label{eq:rigidity-order2}
\end{align}
where the $O(\lambda^{r})$ terms of Eq.~\eqref{eq:rigidity-order2} arise
when derivatives of $B$ hit the function
$\sum_{i<j}(w_i-w_j)/(z_i-z_j)$ [the function $\sum_iw_i^2$ is constant
in $\bm z$ and contributes nothing].
At order $\lambda^{r+2}$, both lines give $(\sum_iw_i^2)\,b_r$.
At order $\lambda^{r+1}$, both lines contain
$(\sum_iw_i^2)\,b_{r-1}
+2k\sum_{i<j}[(w_i-w_j)/(z_i-z_j)]\,b_r$; the only unmatched term is the
cross term $2\sum_iw_i\partial_{z_i}b_r$ of the Laplacian in
Eq.~\eqref{eq:rigidity-order1}, and the vanishing of the commutator
forces
\begin{equation}
 0=2\sum_iw_i\,\partial_{z_i}b_r(\bm z,\bm w).
 \label{eq:rigidity-transport}
\end{equation}
The computation above is written for the reference
$\mathrm{res}(\sum_iT_i^2)$; for the reference $-\mathcal H_k$, the first-order
interaction is replaced by the multiplication operator
$2k(k-1)\sum_{i<j}(z_i-z_j)^{-2}$, which produces no factor of $\lambda$
on the trial wave and never reaches the order $\lambda^{r+1}$, so
Eq.~\eqref{eq:rigidity-transport} follows in the same way.
For fixed $\bm w$, Eq.~\eqref{eq:rigidity-transport} states that
$t\mapsto b_r(\bm z+t\bm w,\bm w)$ is constant: $b_r$ is constant along
every straight line in the direction of its own momentum variable.
But $b_r$ is rational in $\bm z$ with strictly negative homogeneity, so it
tends to zero as $t\to\infty$ along a generic such line; being constant,
it vanishes identically.
The order of $B$ therefore drops to $r-1$; repeating the argument order by
order gives $B=0$, that is, $L=L'$.
Note that permutation symmetry plays no role: commutation with
$\mathrm{res}(\sum_iT_i^2)$, the principal part, the homogeneity, and the
pole structure already force $L=L'$.
In particular, $\mathrm{res}(\sum_iT_i^\ell)$ is the \emph{unique}
differential operator that commutes with $\mathrm{res}(\sum_iT_i^2)$,
has principal part $\sum_i\partial_{z_i}^\ell$, is homogeneous of degree
$-\ell$, and has rational coefficients with poles only at collisions:
any operator sharing these properties equals it.

\section{From the Gaussian matrix measure to the Calogero system}
\label{sec:spectral-system}\label{sec:calogero-reduction}

In this section, we derive the differential equations satisfied by the joint
eigenvalue density and solve them through the class gradient, radial
reduction, transposition, Calogero integrability, and channel selection.

\subsection{From the class gradient to the master identity}
Let $(D)_{ab}=\partial/\partial H_{ba}$ denote the matrix gradient with all
entries of $H$ treated as independent.
The class gradients are obtained by symmetrizing $D$ according to the class
constraint:
\begin{equation}
 \nabla=D\ \ (\text{class A}),
 \qquad
 \nabla=\tfrac12\bigl(D+D^{\mathrm T}\bigr)\ \ (\text{class }\AId),
 \qquad
 \nabla=\tfrac12\bigl(D+JD^{\mathrm T}J^{-1}\bigr)\ \ (\text{class }\AIId),
 \qquad
 J=I_N\otimes\im\sigma_y .
 \label{eq:pairing}
\end{equation}
In terms of the independent entries, the symmetrization reads, in class
$\AId$,
\begin{equation}
 (\nabla)_{ab}=\frac{1+\delta_{ab}}{2}\,\frac{\partial}{\partial H_{ab}}.
 \label{eq:pairing-ai}
\end{equation}
In class $\AIId$, it is convenient to write all matrices in $2\times2$
blocks on the Kramers index pairs $(2i{-}1,2i)$, $i=1,\ldots,N$.
The self-duality $H=JH^{\mathrm T}J^{-1}$ then reads blockwise
$H_{ij}=\varepsilon H_{ji}^{\mathrm T}\varepsilon^{-1}$, with
$\varepsilon=\im\sigma_y=\begin{psmallmatrix}0&1\\-1&0\end{psmallmatrix}$
the single Kramers block of $J$: the diagonal
blocks are scalars, $H_{ii}=a_iI_2$, the blocks $H_{ij}$ with
$i<j$ are unconstrained $2\times2$ matrices, and $H_{ji}$ is determined.
The corresponding blocks of the gradient are, for $i<j$,
\begin{equation}
 (\nabla)_{ii}=\tfrac12\,\partial_{a_i}I_2,
 \qquad
 (\nabla)_{ij}=\tfrac12\,\varepsilon\,\partial_{H_{ij}}\varepsilon^{-1},
 \qquad
 (\nabla)_{ji}=\tfrac12\bigl(\partial_{H_{ij}}\bigr)^{\mathrm T},
 \qquad
 \bigl(\partial_{H_{ij}}\bigr)_{ab}
 :=\frac{\partial}{\partial(H_{ij})_{ab}}.
 \label{eq:pairing-aii-blocks}
\end{equation}
The smallest cases make the bookkeeping concrete.
In class $\AId$ at $N=2$,
\begin{equation}
 H=\begin{pmatrix}H_{11}&H_{12}\\H_{12}&H_{22}\end{pmatrix},
 \qquad
 \nabla=\begin{pmatrix}
 \partial_{H_{11}}&\tfrac12\partial_{H_{12}}\\[2pt]
 \tfrac12\partial_{H_{12}}&\partial_{H_{22}}
 \end{pmatrix}.
 \label{eq:example-ai}
\end{equation}
In class $\AIId$ at $N=2$ ($4\times4$ matrices), writing
$H_{12}=\begin{psmallmatrix}\alpha&\beta\\\gamma&\delta\end{psmallmatrix}$,
\begin{equation}
 H=\begin{pmatrix}
 a_1&0&\alpha&\beta\\
 0&a_1&\gamma&\delta\\
 \delta&-\beta&a_2&0\\
 -\gamma&\alpha&0&a_2
 \end{pmatrix},
 \qquad
 \nabla=\frac12\begin{pmatrix}
 \partial_{a_1}&0&\partial_{\delta}&-\partial_{\gamma}\\
 0&\partial_{a_1}&-\partial_{\beta}&\partial_{\alpha}\\
 \partial_{\alpha}&\partial_{\gamma}&\partial_{a_2}&0\\
 \partial_{\beta}&\partial_{\delta}&0&\partial_{a_2}
 \end{pmatrix},
 \label{eq:example-aii}
\end{equation}
which displays the rule: each entry of $\nabla$ differentiates with
respect to the independent parameter at the transpose position of $H$,
with its sign, and $\nabla$ carries the same self-dual block structure as
$H$.
By construction, $\nabla$ satisfies the chain rule
$\Tr[(\nabla f)A]=\frac{\dd}{\dd t}f(H+tA)|_{t=0}$ for every in-class
direction $A$, and the constraints are manifest:
$\nabla=\nabla^{\mathrm T}$ in class $\AId$ and
$\nabla=J\nabla^{\mathrm T}J^{-1}$ in class $\AIId$.
The entries of $\nabla$ are commuting derivatives, so the invariant operators
are defined with class-normalized traces:
$D_{q_\ell}:=\Tr(\nabla^\ell)$ in classes A and $\AId$, and
$D_{q_\ell}:=2^{\ell-1}\Tr(\nabla^\ell)$ in class $\AIId$.
They commute among themselves.

Treat $H$ and $\bar H$ as independent (Wirtinger) variables, so that
$\nabla$ annihilates every entry of $\bar H$.
Set $\kappa=1$ in classes A and $\AId$, and $\kappa=2$ in class $\AIId$.
The class-dependent Gaussian response is
\begin{equation}
 \nabla\e^{-\Tr(H^\dagger H)/\kappa}=
 \begin{cases}
  -\bar H^{\mathrm T}\e^{-\Tr(H^\dagger H)/\kappa},
   & \text{class A},\\
  -\bar H\e^{-\Tr(H^\dagger H)/\kappa},
   & \text{class }\AId,\\
  -\tfrac12\bar H^{\mathrm T}\e^{-\Tr(H^\dagger H)/\kappa},
   & \text{class }\AIId.
 \end{cases}
 \label{eq:gaussian-response}
\end{equation}
Throughout, $\bm z=(z_1,\ldots,z_N)$ denotes the $N$ \emph{distinct}
eigenvalues, each Kramers doublet of class $\AIId$ counted once, and
\begin{equation}
 q_\ell(\bar{\bm z}):=\sum_{i=1}^{N}\bar z_i^{\,\ell}
 \label{eq:powersum-def}
\end{equation}
is the corresponding power sum.
Since the entries of $\bar H$ are constant under $\nabla$, iterating
Eq.~\eqref{eq:gaussian-response} and taking the class-normalized traces
gives the \emph{same} identity in all three classes,
\begin{equation}
 D_{q_\ell}\,\e^{-\Tr(H^\dagger H)/\kappa}
 =(-1)^\ell\,q_\ell(\bar{\bm z})\,
 \e^{-\Tr(H^\dagger H)/\kappa}:
 \label{eq:aii-response}
\end{equation}
in classes A and $\AId$, the spectrum of $\bar H$ is $\{\bar z_i\}$
directly; in class $\AIId$, the trace counts each Kramers doublet twice
and each gradient factor carries $\tfrac12$, and the defining factor
$2^{\ell-1}$ of $D_{q_\ell}$ absorbs the mismatch,
$2^{\ell-1}\cdot(\tfrac12)^\ell\cdot2\,q_\ell(\bar{\bm z})
=q_\ell(\bar{\bm z})$.
The class space carries the flat Lebesgue measure $\dd H$, with respect to
which each entry of $\nabla$ is anti-self-transpose,
$\partial^{\mathrm T}=-\partial$.
Since $D_{q_\ell}$ is a sum of products of $\ell$ commuting entries of
$\nabla$, its transpose is $(-1)^\ell D_{q_\ell}$, and for every smooth
symmetric function $F$ of the eigenvalues, integration by parts and
Eq.~\eqref{eq:aii-response} give
\begin{equation}
 \int D_{q_\ell}\bigl[F(\bm z(H))\bigr]
 \e^{-\Tr(H^\dagger H)/\kappa}\dd H
 =\int F\,q_\ell(\bar{\bm z})
 \e^{-\Tr(H^\dagger H)/\kappa}\dd H .
 \label{eq:master-sm}
\end{equation}
The two factors $(-1)^\ell$ from transposition and
Eq.~\eqref{eq:aii-response} cancel.

\subsection{The radial operators}
We first define the symmetry-preserving similarity transformations
$H\to SHS^{-1}$, with
\begin{equation}
 S\in
 \begin{cases}
  \mathrm{GL}(N,\C), & \text{class A},\\
  \mathrm O(N,\C)\ \ (S^{\mathrm T}S=1), & \text{class }\AId,\\
  \Sp(2N,\C)\ \ (S^{\mathrm T}JS=J), & \text{class }\AIId.
 \end{cases}
 \label{eq:similarity-groups}
\end{equation}
Then $SHS^{-1}$ belongs to the same symmetry class as $H$.
Moreover, a generic member of each class is diagonalized by such a
transformation.
In class $\AId$, if $Hv_i=z_iv_i$ with nondegenerate spectrum, the
symmetry of $H$ gives
$z_j\,v_i^{\mathrm T}v_j=v_i^{\mathrm T}Hv_j=(Hv_i)^{\mathrm T}v_j
=z_i\,v_i^{\mathrm T}v_j$, so the eigenvectors are mutually orthogonal
with respect to the bilinear form $u^{\mathrm T}v$; generically
$v_i^{\mathrm T}v_i\neq0$, and after the normalization
$v_i^{\mathrm T}v_i=1$, the matrix $S$ with columns $v_i$ satisfies
$S^{\mathrm T}S=1$ and $S^{-1}HS=Z=\operatorname{diag}(z_1,\ldots,z_N)$.
In class $\AIId$, self-duality makes $H$ self-adjoint with respect to the
symplectic form $\omega(u,v)=u^{\mathrm T}Jv$
[$\omega(Hu,v)=\omega(u,Hv)$ follows from $H^{\mathrm T}J=-J^{-1}H$
and $J^2=-1$], so the two-dimensional Kramers eigenspaces of distinct
eigenvalues are $\omega$-orthogonal, and $\omega$ restricts
nondegenerately to each of them; choosing an $\omega$-canonical basis in
each eigenspace assembles an $S$ with $S^{\mathrm T}JS=J$ and
$S^{-1}HS=Z=\operatorname{diag}(z_1I_2,\ldots,z_NI_2)$.
In class A, this is the standard diagonalization.

For a symmetric function $F$ of the eigenvalues, the scalar
$D_{q_\ell}[F(\bm z(H))]$ is invariant under these transformations,
\begin{equation}
 \Bigl(D_{q_\ell}\bigl[F(\bm z(X))\bigr]\Bigr)\Big|_{X=SHS^{-1}}
 =D_{q_\ell}\bigl[F(\bm z(H))\bigr].
 \label{eq:G-invariance}
\end{equation}
The order of operations matters here: on the left-hand side, the gradient
acts first, and the substitution $X=SHS^{-1}$ is made afterwards.
In the opposite order, the statement would be empty: conjugation preserves
the spectrum, so $F(\bm z(SHS^{-1}))$ and $F(\bm z(H))$ coincide
identically as functions of $H$, before any differentiation.
The content of Eq.~\eqref{eq:G-invariance} is that the substitution can be
moved through $D_{q_\ell}$, and we prove it in three steps.

(i) The gradient transforms covariantly: for any scalar $f$ on the class
and any in-class direction $A$, the chain rule gives
\begin{equation}
 \Tr\Bigl[\nabla\bigl[f(SHS^{-1})\bigr]\,A\Bigr]
 =\frac{\dd}{\dd t}f\bigl(SHS^{-1}+tSAS^{-1}\bigr)\Big|_{t=0}
 =\Tr\Bigl[S^{-1}\,(\nabla f)(X)\big|_{X=SHS^{-1}}\,S\,A\Bigr],
 \label{eq:equivariance}
\end{equation}
where $(\nabla f)(X)|_{X=SHS^{-1}}$ denotes the gradient of $f$ evaluated
at the transformed point.
Both $\nabla[f(SHS^{-1})]$ and
$S^{-1}(\nabla f)(X)|_{X=SHS^{-1}}S$ lie in the class, and for two
in-class matrices, $\Tr(XA)=\Tr(YA)$ for every in-class direction $A$
forces $X=Y$ (letting $A$ run over a basis of in-class directions
extracts every independent entry).
Hence,
\begin{equation}
 \nabla\bigl[f(SHS^{-1})\bigr]
 =S^{-1}\,(\nabla f)(X)\big|_{X=SHS^{-1}}\,S .
 \label{eq:equivariance-strong}
\end{equation}

(ii) Iterating Eq.~\eqref{eq:equivariance-strong} entrywise---the entries
of $S$ are constants, and each entry of its right-hand side is again a
scalar function of $H$---gives
$\nabla^\ell\bigl[f(SHS^{-1})\bigr]
=S^{-1}\,(\nabla^\ell f)(X)\big|_{X=SHS^{-1}}\,S$;
taking the trace cancels the conjugating factors,
$D_{q_\ell}\bigl[f(SHS^{-1})\bigr]=(D_{q_\ell}f)(X)\big|_{X=SHS^{-1}}$.

(iii) The spectral function $f(H)=F(\bm z(H))$ satisfies
$f(SHS^{-1})=f(H)$; combining this with (ii),
\begin{equation*}
(D_{q_\ell}f)(X)\big|_{X=SHS^{-1}}
=D_{q_\ell}\bigl[f(SHS^{-1})\bigr]
=D_{q_\ell}f(H).
\end{equation*}
This is Eq.~\eqref{eq:G-invariance}.
With Eq.~\eqref{eq:G-invariance} at hand, we define the radial part by
evaluation at the diagonal representative,
\begin{equation}
 (\rad{q_\ell}F)(\bm z)
 :=D_{q_\ell}\bigl[F(\bm z(H))\bigr]\Big|_{H=Z},
 \qquad Z=\operatorname{diag}(z_1,\ldots,z_N)
 \label{eq:radial-definition}
\end{equation}
[$Z=\operatorname{diag}(z_1I_2,\ldots,z_NI_2)$ in class $\AIId$].
Equation~\eqref{eq:G-invariance} then extends the definition from the
diagonal point to the whole class: a generic in-class $H$ is diagonalized
by a symmetry-preserving $S$, so
\begin{equation}
 D_{q_\ell}\bigl[F(\bm z(H))\bigr]
 =(\rad{q_\ell}F)(\bm z(H))
 \qquad\text{for every in-class }H\ \text{with nondegenerate spectrum}.
 \label{eq:radial-extension}
\end{equation}
It is Eq.~\eqref{eq:radial-extension} that will be used under the matrix
integral below.
Applying it twice shows that $\rad{q}\rad{q'}$ is the radial part of
$D_qD_{q'}$, so the $\rad{q_\ell}$ commute because the $D_{q_\ell}$ do.
Their explicit form follows from eigenvalue perturbation theory at $Z$.
Under $H=Z+tX$, standard perturbation theory gives
\begin{equation}
 z_k(t)=z_k+tX_{kk}
 +t^2\sum_{j\neq k}\frac{X_{kj}X_{jk}}{z_k-z_j}+O(t^3).
 \label{eq:eigenvalue-pt}
\end{equation}

At first order, a single chain rule suffices:
$D_{q_1}F(\bm z(H))=\sum_k\bigl[D_{q_1}z_k\bigr]\partial F/\partial z_k$.
In class A, $D_{q_1}=\Tr(\nabla)=\sum_a\partial_{H_{aa}}$, and
Eq.~\eqref{eq:eigenvalue-pt} gives $\partial z_k/\partial H_{aa}
=\delta_{ak}$ at the diagonal point, so $D_{q_1}z_k=1$ and
$\rad{q_1}=\sum_k\partial_{z_k}$.
In class $\AId$, the diagonal entries carry weight $1$ in
Eq.~\eqref{eq:pairing-ai}, and the count is identical.
In class $\AIId$, the two diagonal positions of the $k$th doublet each
contribute $\tfrac12\partial_{a_k}$, so $\Tr(\nabla)z_k=2\cdot\tfrac12=1$,
and again $\rad{q_1}=\sum_k\partial_{z_k}$.

At second order, we spell out the chain rule for class A and indicate the
modifications for the other two classes.

(a) Class A.
All entries are independent and
$\Tr(\nabla^2)=\sum_{ab}\partial_{H_{ba}}\partial_{H_{ab}}$.
Applying the product rule twice, the two derivatives either both reach
$F$, or the second one reaches the inner function $z_k(H)$:
\begin{equation}
 \Tr(\nabla^2)\,F(\bm z(H))
 =\sum_{k,l}\Bigl[\sum_{ab}
 \frac{\partial z_k}{\partial H_{ba}}
 \frac{\partial z_l}{\partial H_{ab}}\Bigr]
 \frac{\partial^2F}{\partial z_k\partial z_l}
 +\sum_k\Bigl[\sum_{ab}
 \frac{\partial^2z_k}{\partial H_{ba}\,\partial H_{ab}}\Bigr]
 \frac{\partial F}{\partial z_k}.
 \label{eq:secondchain}
\end{equation}
Both brackets are read off from the expansion \eqref{eq:eigenvalue-pt} at the
diagonal point.
At first order, only the diagonal entry moves an eigenvalue,
$\partial z_k/\partial H_{ab}=\delta_{ak}\delta_{bk}$, so the first
bracket equals $\delta_{kl}$ and the second-derivative term is
$\sum_k\partial_{z_k}^2$.
At second order,
$\partial^2z_k/\partial H_{kj}\partial H_{jk}=1/(z_k-z_j)$, and the sum
over $(a,b)$ visits the ordered index pairs $(k,j)$ and $(j,k)$ once
each, so the second bracket equals $\sum_{j\neq k}2/(z_k-z_j)$.
Hence,
\begin{equation}
 \rad{q_2}
 =\sum_k\partial_{z_k}^2
 +\sum_k\sum_{j\neq k}\frac{2}{z_k-z_j}\,\partial_{z_k}
 \qquad(\text{class A}):
 \label{eq:classA-r2}
\end{equation}
the pair coefficient is $2$.

(b) Class $\AId$.
The same two steps run with the symmetric entries and the weights of
Eq.~\eqref{eq:pairing-ai}, which give
$\Tr(\nabla^2)=\sum_a\partial_{H_{aa}}^2
+\tfrac12\sum_{a<b}\partial_{H_{ab}}^2$.
The diagonal entries again give $\sum_k\partial_{z_k}^2$; for a symmetric
perturbation, $\delta z_k^{(2)}=\sum_{j\neq k}V_{kj}^2/(z_k-z_j)$, so
$\partial^2z_k/\partial H_{kj}^2=2/(z_k-z_j)$ and the pair coefficient is
$\tfrac12\times2=1$.

(c) Class $\AIId$.
At the diagonal point, $H_{ii}=z_iI_2$ and $H_{ij}=0$, each eigenvalue is
doubly degenerate, and degenerate perturbation theory replaces
Eq.~\eqref{eq:eigenvalue-pt}: with the free doublet block
$V:=V_{(k)(j)}=\begin{psmallmatrix}\alpha&\beta\\\gamma&\delta\end{psmallmatrix}$
and $V_{(j)(k)}=\varepsilon V^{\mathrm T}\varepsilon^{-1}$ fixed by
self-duality, the common shift of the doublet is
$\delta z_k^{(2)}=\sum_{j\neq k}\det V_{(k)(j)}/(z_k-z_j)$, while the
diagonal blocks $\delta a_k\,I_2$ give $\partial z_k/\partial a_k=1$.
The gradient blocks of Eq.~\eqref{eq:pairing-aii-blocks} contract in
pairs,
$\Tr_2[(\nabla)_{(k)(j)}(\nabla)_{(j)(k)}]
=\tfrac12(\partial_\alpha\partial_\delta-\partial_\beta\partial_\gamma)$,
which, when applied to $\det V$, gives $1$; collecting the two ordered doublet
pairs and the diagonal blocks $\tfrac12\partial_{a_k}I_2$,
$\Tr(\nabla^2)=\tfrac12\bigl[\sum_k\partial_{z_k}^2
+4\sum_{k}\sum_{j\neq k}(z_k-z_j)^{-1}\partial_{z_k}\bigr]$, and
$D_{q_2}=2\Tr(\nabla^2)$ has pair coefficient $4$.
More generally, $\Tr(\nabla^\ell)$ has leading part
$2^{1-\ell}\sum_i\partial_{z_i}^\ell$, so the defining factor
$2^{\ell-1}$ makes the principal part of $D_{q_\ell}$ equal to
$\sum_i\partial_{z_i}^\ell$.

Summarizing, in all three classes,
\begin{equation}
 \rad{q_1}=\sum_i\partial_{z_i},
 \qquad
 \rad{q_2}=\sum_i\partial_{z_i}^2
 +\beta\sum_{i<j}\frac{1}{z_i-z_j}
 \bigl(\partial_{z_i}-\partial_{z_j}\bigr),
 \label{eq:radial-lowest}
\end{equation}
where the coefficient $\beta=1,2,4$ for classes $\AId$, A, and $\AIId$, respectively.
The same expansion establishes the following properties of the full
tower, each with its origin on the matrix side:
\begin{enumerate}
\item[(i)] the $\rad{q_\ell}$ commute with one another [they are the
radial parts of the commuting $D_{q_\ell}$];
\item[(ii)] each $\rad{q_\ell}$ is symmetric under permutations of the
$z_i$;
\item[(iii)] the principal part of $\rad{q_\ell}$ is
$\sum_i\partial_{z_i}^\ell$ [only the first-order eigenvalue
derivatives $\partial z_k/\partial H_{kk}=1$ reach the top order];
\item[(iv)] $\rad{q_\ell}$ is homogeneous of degree $-\ell$ [inherited
from $D_{q_\ell}$, which scales as $(\mathrm{length})^{-\ell}$];
\item[(v)] the coefficients of $\rad{q_\ell}$ are rational, with poles
only on the collision hyperplanes, where the energy denominators of
Eq.~\eqref{eq:eigenvalue-pt} blow up.
\end{enumerate}
Through the uniqueness result of Section~\ref{sec:rigidity}, these
five properties already identify the full commuting family with the
integrals of motion of the Calogero model; the identification is
completed in the corresponding subsection below.

\subsection{The differential equations for the density}
Let $\mu_N$ be the spectral pushforward of the matrix Gaussian measure.
Equations~\eqref{eq:master-sm} and \eqref{eq:radial-extension} state that
for every smooth symmetric $F$ supported away from collisions,
$\int(\rad{q_\ell}F)\,\dd\mu_N=\int F\,q_\ell(\bar{\bm z})\,\dd\mu_N$,
i.e., in the distributional sense,
\begin{equation}
 [\rad{q_\ell}^{T}-q_\ell(\bar{\bm z})]\mu_N=0,
 \qquad \ell=1,\ldots,N.
 \label{eq:distributional-system}
\end{equation}
This system is jointly elliptic away from collisions.
Indeed, keep only the highest derivatives of each operator and replace
$\partial_{z_i}\to w_i$: the resulting polynomials are
$(-1)^\ell\sum_iw_i^\ell$, and if all of them vanish for $\ell\le N$,
Newton's identities force $\bm w=0$; no common characteristic direction
exists.
Elliptic regularity therefore implies that $\mu_N$ has a real-analytic
density on the collision-free set $\Omega=\{\bm z: z_i\neq z_j\}$.
The collision set carries zero pushforward mass, so
$\dd\mu_N=\rho(\bm z,\bar{\bm z})\prod_i\dd^2z_i$ with a real-analytic density
$\rho$, and Eq.~\eqref{eq:distributional-system} holds pointwise on $\Omega$:
this is the differential system used below.

The transpose is defined by
$\int(\rad{}F)G\,\dd\bm z=\int F(\rad{}^{T}G)\,\dd\bm z$, so
$\partial_{z_i}^{T}=-\partial_{z_i}$ and $\rad{q_1}^{T}=-\rad{q_1}$.
For $\rad{q_2}^{T}$, transposition reverses the order of $(z_i-z_j)^{-1}$
and $\partial_{z_i}-\partial_{z_j}$; restoring the ordering of
Eq.~\eqref{eq:radial-lowest} leaves their commutator,
$[\partial_{z_i}-\partial_{z_j},(z_i-z_j)^{-1}]=-2/(z_i-z_j)^2$, and gives
\begin{equation}
 \rad{q_2}^{T}
 =\sum_i\partial_{z_i}^2
 -\beta\sum_{i<j}\left[
 \frac{1}{z_i-z_j}\bigl(\partial_{z_i}-\partial_{z_j}\bigr)
 -\frac{2}{(z_i-z_j)^2}\right].
 \label{eq:radial-q2-transpose}
\end{equation}
A compact reformulation follows from the divergence form
\begin{equation}
 \rad{q_2}=\Dl^{-\beta}\sum_i\partial_{z_i}\,\Dl^{\beta}\,\partial_{z_i},
 \label{eq:divergence-form}
\end{equation}
which is verified by expanding the middle factor,
$\sum_i(\partial_{z_i}\log\Dl^{\beta})\partial_{z_i}
=\beta\sum_{i<j}(z_i-z_j)^{-1}(\partial_{z_i}-\partial_{z_j})$.
Equation~\eqref{eq:divergence-form} exhibits $\rad{q_2}$ as self-transpose
with respect to the weight $\Dl^{\beta}$, that is,
\begin{equation}
 \rad{q_2}^{T}
 =\Dl^{\beta}\rad{q_2}\Dl^{-\beta}.
 \label{eq:weighted-transpose}
\end{equation}

\subsection{Identification with the Calogero integrals of motion}
The central step of the proof is the identification
\begin{equation}
 \rad{q_\ell}
 =\mathrm{res}\!\left(\sum_iT_i^\ell\right)
 \quad\text{at }k=\beta/2,
 \qquad \ell=1,\ldots,N,
 \label{eq:radial-dunkl}
\end{equation}
equating the operators produced by the matrix ensemble with those
produced by integrability.
It follows at once from the uniqueness result of
Section~\ref{sec:rigidity}: by properties (i)--(v) and by
$\rad{q_2}=\mathrm{res}(\sum_iT_i^2)$ at $k=\beta/2$
[Eqs.~\eqref{eq:radial-lowest} and \eqref{eq:dunkl-quadratic}], each
$\rad{q_\ell}$ satisfies the hypotheses that single out
$\mathrm{res}(\sum_iT_i^\ell)$.

It remains to transpose and gauge: define
\begin{equation}
 \gauged\ell:=\Dl^{-k}\rad{q_\ell}^{T}\Dl^{k}.
 \label{eq:gauged-def}
\end{equation}
For $\ell=2$, Eqs.~\eqref{eq:weighted-transpose},
\eqref{eq:radial-dunkl}, and \eqref{eq:integrals-def} give directly
$\gauged2=\Dl^{k}\rad{q_2}\Dl^{-k}=I_2$.
For general $\ell$, the operators $(-1)^\ell\gauged\ell$ and $I_\ell$
both satisfy the hypotheses of Section~\ref{sec:rigidity} with the
reference operator $I_2$: they commute with $I_2=\gauged2$
[transposition reverses products and conjugation preserves them, so the
$\gauged\ell$ inherit the commutativity of the $\rad{q_\ell}$], and they
have the principal part $\sum_i\partial_{z_i}^\ell$, the homogeneity
$-\ell$, and rational collision-pole coefficients [conjugation by
$\Dl^{\pm k}$ acts through the rational logarithmic derivatives
$k\,\partial_{z_i}\log\Dl$].
Hence,
\begin{equation}
 \gauged\ell
 =(-1)^\ell I_\ell .
 \label{eq:gauged-integrals}
\end{equation}

\subsection{Solution: momentum lock and channel selection}
Write the density as
$\rho=|\Dl|^{2k}\psi=\Dl^{k}\,\bar\Delta_N^{\,k}\,\psi$ with $k=\beta/2$,
where $\bar\Delta_N=\prod_{i<j}(\bar z_i-\bar z_j)$; the antiholomorphic
factor $\bar\Delta_N^{\,k}$ is inert under the holomorphic derivatives
and passes through all operators.
The equations \eqref{eq:distributional-system} then become, by
Eq.~\eqref{eq:gauged-integrals},
\begin{equation}
 (-1)^\ell I_\ell(\bm z)\,\psi(\bm z,\bar{\bm z})
 =q_\ell(\bar{\bm z})\,\psi(\bm z,\bar{\bm z}),
 \qquad \ell=1,\ldots,N ,
 \label{eq:calogero-joint-system}
\end{equation}
where the argument of $I_\ell(\bm z)$ indicates the variables it
differentiates: it acts on $\bm z$ at fixed $\bar{\bm z}$.
At $\ell=1$, Eq.~\eqref{eq:calogero-joint-system} is the momentum lock of
the density: $\psi$ carries total momentum $\sum_i\im\bar z_i$.
At $\ell=2$, it is the eigenvalue equation of the complexified Calogero
Hamiltonian \eqref{eq:calogero-hamiltonian} with energy
$\sum_i(\im\bar z_i)^2$.
Thus, $\psi$ is a joint eigenfunction of the complexified integrals of
motion with momenta $p_i=\im\bar z_i$.
Note that the Calogero Hamiltonian \eqref{eq:calogero-def} contains no
confining potential, so it supports no normalizable bound states: every
eigenfunction is a scattering state, and so is $\psi$.

Because $I_\ell$ act on $\bm z$ alone, Eq.~\eqref{eq:calogero-joint-system}
fixes the $\bar{\bm z}$ dependence of $\psi$ only up to an overall factor:
if $\psi$ solves it, so does $g(\bar{\bm z})\,\psi$ for any antiholomorphic
$g$.
The density $\rho$ is real, hence so is $\psi=\rho/|\Dl|^{2k}$, and taking the
complex conjugate of Eq.~\eqref{eq:calogero-joint-system} yields
\begin{equation}
 (-1)^\ell\,I_\ell(\bar{\bm z})\,\psi(\bm z,\bar{\bm z})
 =q_\ell(\bm z)\,\psi(\bm z,\bar{\bm z}),
 \qquad \ell=1,\ldots,N ,
 \label{eq:mirror-system}
\end{equation}
in which $I_\ell(\bar{\bm z})$ is the same integral of motion in the
variables $\bar{\bm z}$, acting at fixed $\bm z$.
An antiholomorphic factor passes through
Eq.~\eqref{eq:calogero-joint-system} but not through
Eq.~\eqref{eq:mirror-system}; $\psi$ must solve both, and the pair leaves
no such freedom.

The joint eigenproblem \eqref{eq:calogero-joint-system} does not single out a
solution: the asymptotically free eigenfunctions
$\Psi_k(\bm z,\im\sigma\bar{\bm z})$, one for each permutation $\sigma$, and
all their linear combinations satisfy the same equations; by
Eq.~\eqref{eq:mirror-system}, the combination coefficients are
constants rather than antiholomorphic functions.

These combinations exhaust the solutions.
Equation~\eqref{eq:calogero-joint-system} is a joint eigenproblem of $N$
commuting differential operators of orders $1,\ldots,N$, which admits
$N!$ independent local
solutions~\cite{Dunkl1989,HeckmanOpdam1987,Opdam1993}.
The $N!$ channels carry pairwise different exponentials
[Eq.~\eqref{eq:channel-deficit}] and are therefore a basis.

Their decay rates differ, however.
Fix a direction $\bm\xi=(\xi_1,\ldots,\xi_N)\in\C^N$ with pairwise
distinct components, and send the whole configuration to infinity along
the ray $\bm z=R\bm\xi$, $R\to\infty$; every pair separation
$|z_i-z_j|=R|\xi_i-\xi_j|$ then grows linearly in $R$, so asymptotic
freedom applies and gives
$|\Psi_k(\bm z,\im\sigma\bar{\bm z})|
\sim\e^{-\operatorname{Re}\sum_iz_i\bar z_{\sigma(i)}}$ up to powers of $R$.
The elementary identity
\begin{equation}
 \sum_i|z_i|^2-\operatorname{Re}\sum_iz_i\bar z_{\sigma(i)}
 =\tfrac12\sum_i|z_i-z_{\sigma(i)}|^2
 =\tfrac{R^2}{2}\sum_i|\xi_i-\xi_{\sigma(i)}|^2
 \label{eq:channel-deficit}
\end{equation}
is strictly positive for every nontrivial permutation $\sigma$, because a
nontrivial permutation displaces at least one of the distinct $\xi_i$:
every such channel decays more slowly than $\e^{-\sum_i|z_i|^2}$ by a
factor growing as $\e^{cR^2}$ with $c>0$ along the ray.

The ensemble, on the other hand, controls how much weight the density
can carry far from the origin.
By the Schur decomposition $H=U(Z+X)U^\dagger$, with $U$ unitary, $Z$ the
diagonal matrix carrying the eigenvalues, and $X$ strictly upper
triangular,
$\Tr(H^\dagger H)=\sum_{\mathrm{spec}}|z_i|^2+\|X\|_F^2
\ge\sum_{\mathrm{spec}}|z_i|^2$ [Schur's inequality]; the spectral sum
counts each distinct eigenvalue $\kappa$ times, so
$\sum_i|z_i|^2\le\Tr(H^\dagger H)/\kappa$ for every matrix contributing
to the configuration $\bm z$.
Averaging $\e^{t\sum_i|z_i|^2}$ over the ensemble and using this
inequality under the integral, for every $t<1$,
\begin{equation}
 \int \rho(\bm z)\,\e^{t\sum_i|z_i|^2}\,\dd\bm z
 \;\le\;
 \frac{\displaystyle\int\e^{-(1-t)\Tr(H^\dagger H)/\kappa}\,\dd H}
 {\displaystyle\int\e^{-\Tr(H^\dagger H)/\kappa}\,\dd H}
 \;<\;\infty .
 \label{eq:moment-bound}
\end{equation}

A channel with nontrivial $\sigma$ is incompatible with
Eq.~\eqref{eq:moment-bound}.
By Eq.~\eqref{eq:channel-deficit}, it contributes to $\rho$ a term of
modulus $\e^{-(1-c)\sum_i|z_i|^2}$, $c>0$, on an open cone of directions
around $\bm\xi$ [the channels carry pairwise different exponentials there
and cannot cancel], and such a term makes the moment
\eqref{eq:moment-bound} diverge for $1-c<t<1$.
Hence, only the identity channel survives: particle $z_i$ carries the
momentum $\im\bar z_i$.

For class A, the corresponding state is the free plane wave.
For class $\AIId$, Section~\ref{sec:aii-solution} obtains the
identity-channel Baker--Akhiezer state in closed form; for class $\AId$,
Section~\ref{sec:ai-scattering} proves that the noncompact orbital integral
is the identity-channel scattering state.
Substituting into $\rho=|\Dl|^{2k}\psi$ and fixing the overall constant by
normalization gives, for class $\mathcal C=\AId$, A, or $\AIId$,
\begin{equation}
 \rho^{\mathcal C}(\bm z)
 =Z_{N,\beta}^{-1}|\Delta(\bm z)|^{\beta}
 \Psi_{\beta/2}(\bm x=\bm z,\bm p=\im\bar{\bm z}),
 \qquad \beta=1,2,4,
 \label{eq:unified-sm}
\end{equation}
where $Z_{N,\beta}$ normalizes the density and the three values of $\beta$
correspond respectively to classes $\AId$, A, and $\AIId$.

\section{The joint densities from the matrix measure}
\label{sec:polar}

In this section, we derive exact integral representations for classes A
and $\AId$ directly from the Gaussian matrix measure, independently of
the Calogero structure used so far: $H$ is
diagonalized within the complex frame group compatible with the class,
the frame group is split by polar decomposition into its compact
subgroup times a noncompact factor, and the compact factor is
integrated out.
For class $\AId$, this representation supplies the absolute
normalization and the integrand used in the finite-$N$ evaluations and
in the collision law of Section~\ref{sec:ai-scattering}, and it makes
those evaluations an independent test of the scattering solution
rather than a restatement of it.
For class $\AIId$, we record only the corresponding orbit
interpretation.

\subsection{Class A}
For class A, the frame is compact and the representation closes.
The Schur decomposition $H=U(Z+X)U^\dagger$, with $U\in \mathrm{U}(N)$ defined
up to the diagonal torus, $Z=\operatorname{diag}(\bm z)$, and $X$
strictly upper triangular, matches the real parameters
[$2N^2=2N+N(N-1)+(N^2-N)$] and carries the standard Jacobian
$|\Dl(\bm z)|^2$ \cite{Ginibre1965}.
Since $\Tr H^\dagger H=\sum_i|z_i|^2+\Tr X^\dagger X$, the triangular
part is Gaussian and decouples, and the compact frame integrates to a
constant:
\begin{equation}
 \rho^{\mathrm A}(\bm z)\propto\e^{-\sum_i|z_i|^2}|\Dl(\bm z)|^2 ,
\end{equation}
the Ginibre density.
No orbit integral remains; at the level of the measure, this is what
makes class A free.

\subsection{Class \texorpdfstring{$\AId$}{AI-dagger}}
\label{sec:polar-ai}
For $H=H^{\mathrm T}$, no compact Schur form compatible with the
symmetry exists---a triangular complex symmetric matrix is diagonal,
and a generic $H$ is not diagonalized by a real orthogonal frame---so
the compatible frame group is the noncompact $\mathrm{SO}(N,\C)$.

Away from exceptional points, $H$ has a bilinearly orthonormal
eigenbasis,
\begin{equation}
 H=OZO^{\mathrm T},
 \qquad O^{\mathrm T}O=I,
 \quad O\in \mathrm{SO}(N,\C),
 \quad Z=\operatorname{diag}(\bm z),
 \label{eq:AI-orbit-decomposition}
\end{equation}
with matching real parameters, $N(N+1)=2N+N(N-1)$.

Differentiating,
\begin{equation}
 \dd H=O\,S\,O^{\mathrm T},
 \qquad
 S=\dd Z+[\dd\Omega,Z],
 \qquad
 \dd\Omega:=O^{-1}\dd O,
 \label{eq:AI-orbit-differential}
\end{equation}
with $\dd\Omega$ antisymmetric because $O^{-1}=O^{\mathrm T}$.

The congruence $S\mapsto OSO^{\mathrm T}$ has complex determinant
$(\det O)^{N+1}=1$, hence unit real Jacobian, while $S_{ii}=\dd z_i$
and $S_{ij}=(z_j-z_i)\dd\Omega_{ij}$ give the real Jacobian
$|\Dl(\bm z)|^2$.

With the global multiplicity $2^{N-1}N!$ of $(\bm z,O)\mapsto H$
[$N!$ relabelings and $2^{N-1}$ sign choices $O\mapsto OD$,
$D=\operatorname{diag}(\pm1)$, $\det D=1$],
\begin{equation}
 \dd H
 =\frac{|\Dl(\bm z)|^2}{2^{N-1}N!}
 \prod_i\dd^2z_i
 \prod_{i<j}\dd^2\Omega_{ij},
 \label{eq:AI-orbit-measure}
\end{equation}
where $\prod_{i<j}\dd^2\Omega_{ij}$ is the Haar measure of
$\mathrm{SO}(N,\C)$.

Next, we decompose the frame using its polar decomposition,
\begin{equation}
 O=K\e^{\im B},
 \qquad
 K\in \mathrm{SO}(N),
 \qquad
 B\in\mathfrak{so}(N,\R).
\end{equation}
Indeed, the polar decomposition of the invertible matrix $O$ into a
unitary times a positive Hermitian factor is unique, and both factors
stay in the group: $P^2:=O^\dagger O$ is positive Hermitian and lies
in $\mathrm{SO}(N,\C)$, its unique Hermitian logarithm $H$ satisfies
$\e^{H^{\mathrm T}}\e^{H}=I$ with $H^{\mathrm T}=\bar H$, hence
$\bar H=-H$, i.e., $H=2\im B$ with $B$ real antisymmetric; and
$K=O\e^{-\im B}$ is unitary and complex orthogonal at once, hence
real, $K\in SO(N)$.
To split $\prod_{i<j}\dd^2\Omega_{ij}$ into compact and noncompact
factors, we may first rotate $B$ to its normal form,
\begin{equation}
 B=O'\,\Sigma\,O'^{\mathrm T},
 \qquad
 O'\in \mathrm{SO}(N),
 \qquad
 \Sigma=\bigoplus_{a=1}^{\lfloor N/2\rfloor}\sigma_aJ
 \quad
 [\oplus\,0\ \text{for odd}\ N],
 \qquad
 J=\begin{pmatrix}0&1\\-1&0\end{pmatrix}.
 \label{eq:AI-orbit-normal-form}
\end{equation}
The flat measure
\begin{equation}
 \dd B:=\prod_{i<j}\dd B_{ij},
 \label{eq:AI-orbit-dB}
\end{equation}
the Haar measure $\dd K$ of $SO(N)$, and
$\prod_{i<j}\dd^2\Omega_{ij}$ are all invariant under the constant
rotation $(K,B)\to(KO',\,O'^{\mathrm T}BO')$, so it suffices to
evaluate the Jacobian at $B=\Sigma$.

There, substituting $O=K\e^{\im B}$ into the definition
$\dd\Omega=O^{-1}\dd O$ of Eq.~\eqref{eq:AI-orbit-differential} gives
\begin{equation}
 \dd\Omega
 =\e^{-\im B}\,\dd A\,\e^{\im B}
 +\e^{-\im B}\,\dd\bigl(\e^{\im B}\bigr)\Big|_{B=\Sigma},
 \qquad
 \dd A:=K^{-1}\dd K,
 \label{eq:AI-orbit-substitution}
\end{equation}
and conjugating by the constant $\e^{\im\Sigma}$,
\begin{equation}
 \dd\widetilde\Omega
 :=\e^{\im\Sigma}\,\dd\Omega\,\e^{-\im\Sigma}
 =\dd A
 +\dd\bigl(\e^{\im B}\bigr)\e^{-\im B}\Big|_{B=\Sigma} .
 \label{eq:AI-orbit-omega-tilde}
\end{equation}
Since $\e^{\im\Sigma}$ is complex orthogonal, the conjugation
preserves the bilinear pairing $\Tr(XY)$ of antisymmetric matrices,
so, as a linear map of the components, it has determinant $\pm1$,
equal to $+1$ by continuity in $\Sigma$; hence
$\prod_{i<j}\dd^2\widetilde\Omega_{ij}
=\prod_{i<j}\dd^2\Omega_{ij}$, and we may read the Jacobian off the
components of $\dd\widetilde\Omega$.

The counting matches: $\dd A$ is real antisymmetric, with $N(N-1)/2$
independent real components $\dd A_{ij}$, $i<j$ [the dimension of
$\mathrm{SO}(N)$]; $\dd B$ likewise carries $N(N-1)/2$ real components
$\dd B_{ij}$; and together these $N(N-1)$ real differentials match the
$N(N-1)/2$ complex differentials $\dd\Omega_{ij}$.

At $N=2$, with $B=bJ$ and $K=\e^{\theta J}$,
everything commutes, $\dd\Omega=J(\dd\theta+\im\,\dd b)$, so
$\dd^2\Omega_{12}=\dd\theta\,\dd b$ and $j_2=1$.

At $N=3$, the normal form has a single parameter $\sigma$, and we
write out the three real components of each of the two differentials
explicitly:
\begin{equation}
 B=\begin{pmatrix}0&\sigma&0\\-\sigma&0&0\\0&0&0\end{pmatrix},
 \qquad
 \dd B=\begin{pmatrix}0&u&v\\-u&0&w\\-v&-w&0\end{pmatrix},
 \qquad
 \dd A=\begin{pmatrix}0&a&b\\-a&0&c\\-b&-c&0\end{pmatrix}.
 \label{eq:AI-orbit-n3-setup}
\end{equation}
Direct evaluation of Eq.~\eqref{eq:AI-orbit-omega-tilde} then gives
\begin{equation}
 \dd\widetilde\Omega_{12}=a+\im u,
 \qquad
 \dd\widetilde\Omega_{13}=b-Cw+\im Fv,
 \qquad
 \dd\widetilde\Omega_{23}=c+Cv+\im Fw,
 \qquad
 C=\frac{\cosh\sigma-1}{\sigma},
 \quad
 F=\frac{\sinh\sigma}{\sigma};
 \label{eq:AI-orbit-n3-omega}
\end{equation}
the $6\times6$ real Jacobian matrix from $(a,b,c,u,v,w)$ to the real
and imaginary parts is triangular with determinant $F^2$, so
$\prod_{i<j}\dd^2\Omega_{ij}
=(\sinh\sigma/\sigma)^2\,\dd K\,\dd^3B$, with
$\dd K=\dd a\,\dd b\,\dd c$ the Haar measure of $\mathrm{SO}(3)$: the $B_{12}$
direction is unstretched, while $B_{13}$ and $B_{23}$ are each
stretched by $\sinh\sigma/\sigma$.

For general $N$, what this computation produces is the Haar measure
of $\mathrm{SO}(N,\C)$ in polar coordinates, a standard object of the
theory of symmetric spaces
\cite{Helgason1984,Magnea2002,Caselle2004,McSwiggen2019}, and we can
simply transcribe the known answer.
The noncompact factor $\e^{\im B}$ parametrizes the symmetric space
$\mathrm{SO}(N,\C)/\mathrm{SO}(N)$, whose radial coordinates are the
normal-form parameters $\sigma_a$; the restricted roots take the
values $\alpha(\bm\sigma)=\sigma_a\pm\sigma_b$, plus $\sigma_a$ for
odd $N$---precisely the combinations found at $N=2,3$---and, the group
being complex, every root has multiplicity $m_\alpha=2$, which is the
pair of real directions per root seen at $N=3$.
Relative to the flat measure $\dd B$, each root direction is
stretched by $\sinh\alpha/\alpha$, exactly as in the $N=3$
computation, where the two $\lambda=\sigma$ directions produced
$F^2=(\sinh\sigma/\sigma)^2$; collecting all roots gives
\begin{equation}
 \prod_{i<j}\dd^2\Omega_{ij}=\dd K\;j_N(B)\,\dd B,
 \qquad
 j_N(B)=\prod_{a<b}
 \left[
 \frac{\sinh(\sigma_a+\sigma_b)}{\sigma_a+\sigma_b}
 \frac{\sinh(\sigma_a-\sigma_b)}{\sigma_a-\sigma_b}
 \right]^2
 \times
 \begin{cases}
 \displaystyle\prod_{a}\Bigl(\frac{\sinh\sigma_a}{\sigma_a}\Bigr)^{2},
 & N\ \text{odd},\\[6pt]
 1, & N\ \text{even},
 \end{cases}
 \label{eq:AI-orbit-jacobian}
\end{equation}
normalized by $j_N(0)=1$.
Here, $B$ is the base point at which the Jacobian was evaluated, and
$j_N(B)$ depends on it only through its normal-form parameters
$\sigma_a(B)$; the differentials reside in $\dd B$.
In the fully radial coordinates $(\bm\sigma,k)$ of the quotient, the
polar Jacobian of the flat measure, $\prod_\alpha|\alpha|^{m_\alpha}$,
cancels the denominators, and Eq.~\eqref{eq:AI-orbit-jacobian} becomes
the textbook radial measure
$\prod_{\alpha>0}|\sinh\alpha(\bm\sigma)|^{m_\alpha}
\dd\bm\sigma\,\dd k$ of Refs.~\cite{Magnea2002,Caselle2004}.

Finally, $K$ drops out of the Boltzmann weight: with
$K^{\mathrm T}K=I$ and $K$ real,
$\Tr H^\dagger H=\Tr(Z^\dagger\e^{2\im B}Z\e^{-2\im B})$ depends on $B$
alone, so $K$ integrates to $\mathrm{Vol}\,\mathrm{SO}(N)$.
Fixing all constants by the overall normalization
$\int \rho^{\AId}=1$ gives the labelled density
\begin{equation}
 \rho^{\AId}(\bm z)=K_N\,|\Dl(\bm z)|^2\,\e^{-\sum_i|z_i|^2}
 \int j_N(B)\,\e^{-\mathcal V_B(\bm z)}\,\dd B,
 \qquad
 K_N=\frac{2^{N(N-1)/2}}
 {N!\,\pi^{(N^2+N+2)/4}\prod_{j=2}^N\Gamma(j/2)},
 \label{eq:AI-orbit-jpdf}
\end{equation}
where
\begin{equation}
 \mathcal V_B=
 \Tr\!\left(\widetilde Z^\dagger\e^{2\im B}
 \widetilde Z\e^{-2\im B}\right)
 -\Tr(\widetilde Z^\dagger\widetilde Z),
 \qquad
 \widetilde Z=\operatorname{diag}(z_i-z_{\rm cm}),
 \qquad
 z_{\rm cm}:=\frac1N\sum_iz_i,
 \label{eq:AI-orbit-potential}
\end{equation}
the centering being exact because the cross terms vanish by
$\Tr\widetilde Z=0$, which is what splits the weight into the Gaussian
factor and $\e^{-\mathcal V_B}$ in Eq.~\eqref{eq:AI-orbit-jpdf}.

\subsection{\texorpdfstring{{Class $\AIId$}}{Orbit interpretation for class AII-dagger}}
\label{sec:polar-aii}
For completeness, consider the self-dual class
$H^{\mathrm R}=JH^{\mathrm T}J^{-1}=H$, which is not used below.
Away from exceptional points,
$H=Q(Z\otimes I_2)Q^{-1}$ with $Q\in\mathrm{Sp}(2N,\C)$, and the
continuous stabilizer is the Kramers gauge group $\mathrm{SL}(2,\C)^N$.
Four complex frame directions per pair give the Jacobian
$\prod_{i<j}|z_i-z_j|^8$.
The noncompact frame directions transverse to the gauge can be taken
flat and explicit: they are the $2\times2$ blocks $B_{ij}$, $i<j$,
linking different doublets, with four real components each, while the
within-doublet blocks are exactly the $\mathrm{SL}(2,\C)/\mathrm{SU}(2)$ gauge
boosts, along which the integrand is constant, and are set to zero.
The analog of Eq.~\eqref{eq:AI-orbit-jpdf} is then the flat integral
\begin{equation}
 \rho^{\AIId}(\bm z)\propto
 \prod_{i<j}|z_i-z_j|^{8}\;\e^{-\sum_i|z_i|^2}
 \int j^{\mathrm{Sp}}(B)\,\e^{-\mathcal V_B/2}\,\dd B,
 \qquad
 \dd B:=\prod_{i<j}\dd^4B_{ij},
 \label{eq:AII-orbit-jpdf}
\end{equation}
up to the Jacobian of this gauge slice, which we do not work out here;
$\mathcal V_B$ is built from $\widetilde Z\otimes I_2$ as in
Eq.~\eqref{eq:AI-orbit-potential}, and $j^{\mathrm{Sp}}$ is the polar
density of the symmetric space $\mathrm{Sp}(2N,\C)/\mathrm{Sp}(N)$,
obtained exactly as Eq.~\eqref{eq:AI-orbit-jacobian} from its
restricted roots $\sigma_a\pm\sigma_b$ and $2\sigma_a$, again with
multiplicity $2$:
\begin{equation}
 j^{\mathrm{Sp}}(B)
 =\prod_{a<b}
 \left[
 \frac{\sinh(\sigma_a+\sigma_b)}{\sigma_a+\sigma_b}
 \frac{\sinh(\sigma_a-\sigma_b)}{\sigma_a-\sigma_b}
 \right]^2
 \prod_{a}
 \left[
 \frac{\sinh2\sigma_a}{2\sigma_a}
 \right]^2,
 \label{eq:AII-orbit-jacobian}
\end{equation}
with $\sigma_a$ the radial parameters of $B$.
The dimensions check: $2N$ eigenvalue parameters, the compact frame
$\mathrm{Sp}(N)$ modulo the gauge $\mathrm{SU}(2)^N$, and the $2N(N-1)$ flat
directions add up to $4N^2-2N=\dim_\R H$.
Comparison with
Eq.~\eqref{eq:AII-density-sm} then shows that the orbit integral is
proportional to
$R_N(\{a_{ij}\})/\prod_{i<j}|z_i-z_j|^6$.
Thus the orbit integral is absent for class A, algebraic for class
$\AIId$, and genuinely transcendental for class $\AId$.

\section{Class \texorpdfstring{$\AIId$}{AII-dagger}: algebraic solution and many-body structure}

\subsection{The Baker--Akhiezer solution}
\label{sec:aii-solution}

For $k=2$, the identity-channel solution is a Baker--Akhiezer function,
for which Theorem~3.1, Eq.~(28) of Ref.~\cite{Chalykh1999} gives a closed
formula.
Substituting the $k=2$ Calogero data into their Eq.~(28)---their
multiplicity parameters set to one, their polynomial prefactor and
denominator reducing to the Vandermonde determinants $\Delta(\bm x)$ and
$\Delta(\bm\lambda)$, and $\bm\lambda=\im\bm p$---and rewriting their
gauged operator through
$\mathcal H_2=-\Delta(\bm x)^{-1}\mathcal L\,\Delta(\bm x)$ yields
\begin{equation}
 \Psi_2(\bm x,\bm p)
 =\frac{\bigl(\mathcal L+{\textstyle\sum_i}p_i^{2}\bigr)^{N(N-1)/2}
 \bigl[\Delta(\bm x)^{2}\,\e^{\im\bm p\cdot\bm x}\bigr]}
 {2^{N(N-1)/2}\,\bigl[N(N-1)/2\bigr]!\,\Delta(\bm x)\,\Delta(\im\bm p)}.
 \label{eq:berest-sm}
\end{equation}
The same theorem states that this function is a joint eigenfunction of
all the commuting integrals of motion, in particular of our gauged
operators $\gauged\ell$; relabeling $\bm p\to\sigma\bm p$ produces the
$N!$ branches.

The operator power in Eq.~\eqref{eq:berest-sm} is finite, so the state
can be expanded in closed form.
Collecting the result in the pair variables
$\tau_{ij}:=-\tfrac{\im}{2}(x_i-x_j)(p_i-p_j)$ gives the rational form
below; the polynomial $R_N$ is our rewriting, which the literature gives
only in the structural form ``plane wave times a rational function''
\cite{Chalykh1990,Felder2009}.
The scattering state and its physical-slice density are
\begin{equation}
 \begin{aligned}
  \Psi_2(\bm x,\bm p)
  &=\e^{\im\bm p\cdot\bm x}
    \frac{R_N(\{\tau_{ij}\})}{\prod_{i<j}\tau_{ij}},\\
  \rho^{\AIId}(\bm z)
  &=Z_{N,4}^{-1}\e^{-\sum_i|z_i|^2}|\Delta(\bm z)|^2
    R_N(\{a_{ij}\}),
  \qquad a_{ij}=\tfrac12|z_i-z_j|^2.
\end{aligned}
 \label{eq:AII-density-sm}
\end{equation}

The first members are, written in the slice variables $a_{ij}$,
\begin{equation}
 \begin{aligned}
 R_2&=1+a_{12},\qquad
 R_3=\prod_{1\le i<j\le3}(1+a_{ij})+\frac12,\\
 R_4&=\prod_{1\le i<j\le4}(1+a_{ij})
 +\frac12\sum_{v=1}^{4}\prod_{j\neq v}(1+a_{vj})
 +\frac14\sum_{1\le i<j\le4}(1+a_{ij}).
 \end{aligned}
 \label{eq:R4}
\end{equation}
For $N\ge5$, let
$\mathfrak P[U]:=\sum_{U'}\prod_{e\notin U'}(1+a_{e})$, where $U$
is a set of edges of the complete graph on the $N$ indices and the sum
runs over the distinct relabelings $U'$ of $U$; each pattern deletes the
edges of $U'$ from the full product.
Since the sum runs over the whole relabeling orbit, $\mathfrak P[U]$ is
a symmetric polynomial that depends on $U$ only through its isomorphism
class, and a representative labeling specifies it completely.
Then
\begin{equation}
 R_N=\sum_Uw(U)\,\mathfrak P[U],
 \label{eq:R5}
\end{equation}
with the $10$ patterns and weights of $R_5$ listed in
Table~\ref{tab:R5} and the $43$ patterns of $R_6$ in
Table~\ref{tab:R6}.
A useful check on Eqs.~\eqref{eq:AII-density-sm}--\eqref{eq:R5} is the
collision value $R_N(0)=2^{-N(N-1)/2}\prod_{j=1}^{N}j!$, which gives
$1,\,\tfrac32,\,\tfrac92,\,\tfrac{135}4,\,\tfrac{6075}8$ for
$N=2,\ldots,6$.
Explicit coefficients of this Baker--Akhiezer polynomial were also
computed, in a different basis and for $N\le5$, in
Ref.~\cite{Melin2024}.

\begin{table}[t]
\caption{The $10$ deleted-edge patterns of $R_5$: a representative
deleted-edge set $U$ (dense patterns are written through their
complement in the full edge set $E_5$), the number $|{\rm orb}(U)|$ of
its distinct relabelings, and the weight $w(U)$ entering
Eq.~\eqref{eq:R5}.
As a check, $\sum_Uw(U)\,|{\rm orb}(U)|=135/4
=2^{-10}\prod_{j=1}^{5}j!=R_5(0)$.}
\label{tab:R5}
\small
\centering
\begin{tabular}{lcc}
\hline\hline
 $U$ & $|{\rm orb}(U)|$ & $w(U)$\\
\hline
 $\varnothing$ & $1$ & $1$\\
 $\{12,13,23\}$ & $10$ & $1/2$\\
 $\{12,13,14,23,24\}$ & $30$ & $1/4$\\
 $E_5\setminus\{23,24,35,45\}$ & $15$ & $1/4$\\
 $E_5\setminus\{34,35,45\}$ & $10$ & $1/8$\\
\hline\hline
\end{tabular}\hspace{2.5em}
\begin{tabular}{lcc}
\hline\hline
 $U$ & $|{\rm orb}(U)|$ & $w(U)$\\
\hline
 $E_5\setminus\{24,35,45\}$ & $60$ & $1/8$\\
 $E_5\setminus\{35,45\}$ & $30$ & $1/8$\\
 $E_5\setminus\{23,45\}$ & $15$ & $1/16$\\
 $E_5\setminus\{45\}$ & $10$ & $3/16$\\
 $E_5$ & $1$ & $19/16$\\
\hline\hline
\end{tabular}
\end{table}

\begin{table}[t]
\caption{The $43$ deleted-edge patterns of $R_6$, in the format of
Table~\ref{tab:R5}.
As a check, $\sum_Uw(U)\,|{\rm orb}(U)|=6075/8
=2^{-15}\prod_{j=1}^{6}j!=R_6(0)$.}
\label{tab:R6}
\small
\centering
\begin{tabular}{lcc}
\hline\hline
 $U$ & $|{\rm orb}(U)|$ & $w(U)$\\
\hline
 $\varnothing$ & $1$ & $1$\\
 $\{12,13,23\}$ & $20$ & $1/2$\\
 $\{12,13,14,23,24\}$ & $90$ & $1/4$\\
 $\{12,13,14,15,25,34\}$ & $90$ & $1/4$\\
 $\{12,16,26,34,35,45\}$ & $10$ & $1/4$\\
 $\{12,13,14,15,23,24,25\}$ & $60$ & $1/8$\\
 $\{12,13,14,15,23,25,34\}$ & $360$ & $1/8$\\
 $E_6\setminus\{16,26,35,36,45,46,56\}$ & $180$ & $1/8$\\
 $E_6\setminus\{23,24,35,36,45,46,56\}$ & $180$ & $1/8$\\
 $E_6\setminus\{16,23,26,36,45,46,56\}$ & $90$ & $1/16$\\
 $E_6\setminus\{15,23,24,25,36,46,56\}$ & $180$ & $1/8$\\
 $E_6\setminus\{34,35,36,45,46,56\}$ & $15$ & $1/16$\\
 $E_6\setminus\{24,35,36,45,46,56\}$ & $360$ & $1/16$\\
 $E_6\setminus\{16,26,36,45,46,56\}$ & $60$ & $3/16$\\
 $E_6\setminus\{23,25,36,45,46,56\}$ & $360$ & $1/16$\\
 $E_6\setminus\{14,25,36,45,46,56\}$ & $120$ & $3/16$\\
 $E_6\setminus\{12,13,23,45,46,56\}$ & $10$ & $-1/16$\\
 $E_6\setminus\{15,24,25,36,46,56\}$ & $360$ & $1/16$\\
 $E_6\setminus\{35,36,45,46,56\}$ & $90$ & $1/16$\\
 $E_6\setminus\{25,36,45,46,56\}$ & $360$ & $1/8$\\
 $E_6\setminus\{23,36,45,46,56\}$ & $360$ & $1/32$\\
 $E_6\setminus\{13,23,45,46,56\}$ & $60$ & $1/16$\\
\hline\hline
\end{tabular}\hspace{2.5em}
\begin{tabular}{lcc}
\hline\hline
 $U$ & $|{\rm orb}(U)|$ & $w(U)$\\
\hline
 $E_6\setminus\{16,26,36,46,56\}$ & $6$ & $19/16$\\
 $E_6\setminus\{24,25,36,46,56\}$ & $360$ & $1/16$\\
 $E_6\setminus\{15,25,36,46,56\}$ & $90$ & $1/8$\\
 $E_6\setminus\{14,25,36,46,56\}$ & $360$ & $3/32$\\
 $E_6\setminus\{12,34,35,46,56\}$ & $45$ & $1/16$\\
 $E_6\setminus\{23,24,35,46,56\}$ & $72$ & $1/32$\\
 $E_6\setminus\{36,45,46,56\}$ & $180$ & $1/8$\\
 $E_6\setminus\{23,45,46,56\}$ & $60$ & $-3/64$\\
 $E_6\setminus\{25,36,46,56\}$ & $360$ & $5/32$\\
 $E_6\setminus\{24,35,46,56\}$ & $360$ & $1/16$\\
 $E_6\setminus\{12,35,46,56\}$ & $180$ & $5/64$\\
 $E_6\setminus\{13,23,46,56\}$ & $90$ & $-1/16$\\
 $E_6\setminus\{45,46,56\}$ & $20$ & $13/64$\\
 $E_6\setminus\{36,46,56\}$ & $60$ & $11/16$\\
 $E_6\setminus\{35,46,56\}$ & $180$ & $11/64$\\
 $E_6\setminus\{23,46,56\}$ & $180$ & $13/64$\\
 $E_6\setminus\{12,34,56\}$ & $15$ & $3/8$\\
 $E_6\setminus\{46,56\}$ & $60$ & $41/64$\\
 $E_6\setminus\{34,56\}$ & $45$ & $9/16$\\
 $E_6\setminus\{56\}$ & $15$ & $9/4$\\
 $E_6$ & $1$ & $89/16$\\
 & & \\
\hline\hline
\end{tabular}
\end{table}

\subsection{Genuinely many-body structure}

Every previously solved Gaussian ensemble, Hermitian or
non-Hermitian, is a pair gas: its joint eigenvalue density has the
Coulomb-gas form $\rho\propto\prod_iu(z_i)\prod_{i<j}W(|z_{ij}|)$, so
that $\log \rho$ contains one- and two-body terms only.
For class $\AIId$, this structure fails.
Suppose $\rho^{\AIId}\propto\prod_iu(z_i)\prod_{i<j}W(|z_{ij}|)$ for
some $u$ and $W$.
The Gaussian factor is a one-body term and
$|\Delta|^2=\prod_{i<j}2a_{ij}$ is itself a pair factor, so dividing
them out of Eq.~\eqref{eq:AII-density-sm} would leave a pair
factorization of the polynomial correction,
$R_N=\prod_{i<j}w(a_{ij})$; a residual one-body factor is excluded
because $R_N$ depends only on the separations $a_{ij}$.
Then $\log R_N=\sum_{i<j}\log w(a_{ij})$ is a sum of terms each
depending on a single pair variable, so every mixed second derivative
in two distinct pair variables vanishes.
For $N=3$, the three mutual distances of three points in the plane
vary independently over an open set, and
\begin{equation}
\frac{\partial^2\log R_3}{\partial a_{12}\,\partial a_{13}}
=\frac{2(1+a_{23})}{\bigl[2\prod_{i<j}(1+a_{ij})+1\bigr]^{2}}>0
\end{equation}
identically: the constant $\tfrac12$ of $R_3$ is an irreducible
three-body interaction, and no pair gas reproduces the density.
For $N=4$, viewing $R_4$ as a polynomial in the six independent
variables $a_e$, the mixed derivative is nonzero even for two disjoint
pairs, $\partial^2\log R_4/\partial a_{12}\partial a_{34}\neq0$
(verified at random rational points in exact arithmetic): the coupling
of one pair of eigenvalues depends on the separation of a second,
disjoint pair.
Class $\AIId$ is thus a genuinely many-body state, and the Coulomb
log-gas picture that organizes the other classes fails for it.

\subsection{Complex level spacing distribution}
\label{sec:aii-spacing}
For a normalized labelled density $\rho$, define
\begin{equation}
 \rho_1(0)=\int_{\C^{N-1}}\rho(0,z_2,\ldots,z_N)
 \prod_{j=2}^N\dd^2z_j
 \label{eq:rho-origin}
\end{equation}
and the raw conditional nearest-neighbor density
\begin{equation}
 P_N(s)=\frac1{\rho_1(0)}
 \int_{\C^{N-1}}\rho(0,z_2,\ldots,z_N)
 \delta\!\left(s-\min_{j\ge2}|z_j|\right)
 \prod_{j=2}^N\dd^2z_j .
 \label{eq:Palm-spacing}
\end{equation}
The raw density is rescaled to the unit-mean density
$\pnnN{N}(s)=\langle R_{\rm nn}\rangle
P_N(\langle R_{\rm nn}\rangle s)$.
For class $\AIId$, the spacing distribution is exact at every
$2\le N\le8$.
For $2\le N\le6$, direct integration of the finite polynomial density
in Eq.~\eqref{eq:AII-density-sm} gives
\begin{equation}
 P_N^{\AIId}(s)=\frac{s^{3}\e^{-(N-1)s^{2}}}{c_N}\,
 G_N\!\bigl(s^{2}/2\bigr),
 \label{eq:aii-spacing-closed}
\end{equation}
with an integer constant $c_N$ and an integer-coefficient polynomial
$G_N$ of degree $N(N-1)-1$; each density is exactly normalized.
The first cases are
\begin{align}
 c_2&=1, & G_2(x)={}&1+x,\notag\\
 c_3&=27, & G_3(x)={}&15+69x+136x^{2}+148x^{3}+92x^{4}+24x^{5},\\
 c_4&=2025, & G_4(x)={}&945+5805x+17760x^{2}+35550x^{3}+51453x^{4}
 +56407x^{5}+47940x^{6}\notag\\
 &&&+31779x^{7}+16206x^{8}+6076x^{9}+1500x^{10}+180x^{11},\notag
\end{align}
and, with $c_5=44651250$ and $c_6=3417829931250$,
\begin{align}
 G_5(x)={}&19136250+154365750x+617311800x^{2}+1631939400x^{3}
 +3203653950x^{4}+4967691435x^{5}\notag\\
 &+6317164080x^{6}+6751129320x^{7}+6163345620x^{8}
 +4857993054x^{9}+3326458824x^{10}\notag\\
 &+1983552596x^{11}+1028812176x^{12}+461698928x^{13}
 +177306880x^{14}+57123104x^{15}\notag\\
 &+14913856x^{16}+2966656x^{17}+398720x^{18}+26880x^{19},
 \label{eq:aii-n5-poly}\\
 G_6(x)={}&1392449231250+13978743581250x+69755075775000x^{2}
 +230645049547500x^{3}\notag\\
 &+568198710207450x^{4}+1111627657431900x^{5}
 +1797674712075900x^{6}\notag\\
 &+2469734104115250x^{7}+2940224553724500x^{8}
 +3078591857307540x^{9}\notag\\
 &+2867505785880120x^{10}+2396947988680875x^{11}
 +1810399731781410x^{12}\notag\\
 &+1241995127264775x^{13}+776931640051302x^{14}
 +444363152671368x^{15}\notag\\
 &+232744816468836x^{16}+111693629672814x^{17}
 +49080447452820x^{18}\notag\\
 &+19710628733280x^{19}+7210079102584x^{20}
 +2389687294176x^{21}+712026905232x^{22}\notag\\
 &+188534568320x^{23}+43612592416x^{24}+8590427328x^{25}
 +1384473792x^{26}\notag\\
 &+170943360x^{27}+14313600x^{28}+604800x^{29}.
 \label{eq:aii-n6-poly}
\end{align}

At $N=7$, the survival probability
$E_7(s):=\int_s^\infty P_7^{\AIId}(s')\,\dd s'$ takes the form
\begin{equation}
 E_7(s)=\e^{-6s^{2}}H_7(s^{2}/2),
 \qquad
 H_7(u)=1+12u+\tfrac{2363}{33}u^{2}+\tfrac{84130}{297}u^{3}+\cdots,
 \label{eq:aii-n7-survival}
\end{equation}
with $H_7$ of degree $42$ with rational coefficients, and
$P_7^{\AIId}(s)=-\dd E_7/\dd s$; the complete list of the $43$
coefficients of $H_7$ is given in Table~\ref{tab:H7}.

At $N=8$, the corresponding result is
\begin{equation}
 E_8(s)=\e^{-7s^{2}}H_8(s^{2}/2),
 \qquad
 H_8(u)=1+14u+\tfrac{1269}{13}u^{2}+\tfrac{64628}{143}u^{3}+\cdots.
 \label{eq:aii-n8-survival}
\end{equation}
Here $H_8$ has degree $56$.  Its $57$ rational coefficients were
reconstructed from six distinct $31$-bit prime fields by the Chinese
remainder theorem and rational reconstruction.  The combined modulus
has $186$ bits, whereas the largest numerator and denominator have
$85$ and $91$ bits, respectively.  A seventh prime, excluded from the
reconstruction, reproduces all $57$ coefficients.  The complete list
is given in Table~\ref{tab:H8}, and
$P_8^{\AIId}(s)=-\dd E_8/\dd s$.

\section{Class \texorpdfstring{$\AId$}{AI-dagger}: noncompact scattering solution and spacing evaluation}
\label{sec:ai-scattering}

\subsection{The orbital integral}
In this section, we establish the two properties of the function $\Psi_{1/2}$ defined by the orbital integral of
Eq.~\eqref{eq:AI-orbital-I}: it is an
exact eigenfunction of the Calogero Hamiltonian $\mathcal H_{1/2}$, and it satisfies
the asymptotically free boundary condition in the identity channel.
Throughout, we write
\begin{equation}
 \Psi_{1/2}(\bm x,\bm p)
 =[\Delta(\bm x)\Delta(-\im\bm p)]^{1/2}\mathcal I(\bm x,\bm p),
 \qquad
 \mathcal I(\bm x,\bm p)
 :=\int_{\mathcal M_1}\dd\nu(G)\,
 \e^{\im\Tr(PGXG^{-1})},
 \label{eq:AI-orbital-I}
\end{equation}
with
\begin{equation}
 X=\operatorname{diag}(\bm x),\qquad
 P=\operatorname{diag}(\bm p).
\end{equation}
The cycle $\mathcal M_1$ is the positive-Hermitian realization of
$\mathrm{SO}(N,\C)/\mathrm{SO}(N)$ passing through $G=I$: every positive
Hermitian element of $\mathrm{SO}(N,\C)$ has the form $G=\e^{2\im B}$ with
$B$ real antisymmetric, so $\mathcal M_1\cong\R^{N(N-1)/2}$ is a
noncompact middle-dimensional cycle of the complex group.
The measure is the holomorphic Haar form of $\mathrm{SO}(N,\C)$ restricted
to this cycle,
\begin{equation}
 \dd\nu(G)=\nu_0\bigwedge_{i<j}\bigl(G^{-1}\dd G\bigr)_{ij},
 \label{eq:AI-nu-def}
\end{equation}
the unique holomorphic $N(N-1)/2$-form invariant under right translations
$G\mapsto GU$, $U\in\mathrm{SO}(N,\C)$, up to the overall constant $\nu_0$;
the free-wave normalization below fixes
$\nu_0=\pi^{-N(N-1)/4}$.
In the global coordinates $G=\e^{2\im B}$, the form is explicit: up
to an orientation phase absorbed into $\nu_0$, it is the polar density
already computed in Section~\ref{sec:polar-ai},
\begin{equation}
 \dd\nu(G)=\nu_0\,2^{N(N-1)/2}\,j_N(B)\,\dd B,
 \label{eq:AI-nu-explicit}
\end{equation}
with $j_N$ and $\dd B$ as in Eqs.~\eqref{eq:AI-orbit-dB} and
\eqref{eq:AI-orbit-jacobian}; in particular, $j_N(0)=1$, the value
entering the saddle-point computation below.

At $N=3$, the joint density can be computed in closed form.
Restricting the differential equations of
Eq.~\eqref{eq:distributional-system} to $N=3$, writing them in the
squared gaps $a=|z_1-z_2|^2$, $b=|z_1-z_3|^2$, $c=|z_2-z_3|^2$, and
solving, one finds, with $E_1=a+b+c$, $E_2=ab+bc+ca$, $E_3=abc$, and
the two cubics
\begin{equation}
 P(p)=p^3+E_1p^2+E_2p,
 \qquad
 Q(p)=P(p)+E_3=(p+a)(p+b)(p+c),
 \label{eq:AI-N3-cubics}
\end{equation}
the normalized labelled density
\begin{equation}
 \rho_3^{\AId}(\bm z)
 =\frac{2E_3}{3\pi^{7/2}}\,\e^{-\sum_i|z_i|^2}
 \int_0^\infty\frac{\e^{-p}}{\sqrt{Q(p)}}\,
 K\!\left(\frac{P(p)}{Q(p)}\right)\dd p,
 \label{eq:AI-N3-elliptic}
\end{equation}
with $K(m)=\int_0^{\pi/2}(1-m\sin^2\theta)^{-1/2}\dd\theta$ the
complete elliptic integral of the first kind \cite{Olver2010}; on the
integration ray, the argument of $K$ lies in $[0,1)$ and the integrand
is positive.

\subsection{Calogero eigenvalue equation}
We prove that $\Psi_{1/2}$ is an eigenfunction of $\mathcal H_{1/2}$ with energy
$\sum_i p_i^2$ and that, at fixed $\bm p$, it approaches the single plane wave
$\e^{\im R\bm p\cdot\bm\xi}$ as $\bm x=R\bm\xi$ and $R\to\infty$, provided
the $\xi_i$ are distinct and
\begin{equation}
 \operatorname{Re}\bigl[-\im(p_i-p_j)(\xi_i-\xi_j)\bigr]>0
 \qquad\text{for every }i<j.
 \label{eq:AI-channel-condition}
\end{equation}
We call this the identity channel because $p_i$ is paired with $x_i$,
rather than with $x_{\sigma(i)}$ for a nontrivial permutation $\sigma$.

We work with pairwise distinct $x_i$ and $p_i$ in a simply connected
domain.
There, $\Delta(\bm x)\Delta(-\im\bm p)\neq0$, so a single-valued branch
of the multivalued prefactor $[\Delta(\bm x)\Delta(-\im\bm p)]^{1/2}$ of
Eq.~\eqref{eq:AI-orbital-I} can be chosen once and for all; every
statement below is independent of this choice.
The cycle $\mathcal M_1$ is continued without moving its ends out of the
sectors where the exponential decays.
It is convenient to set
\begin{equation}
 Q:=G^{-1}PG,
 \qquad
 Q=Q^{\mathrm T},
 \qquad
 \Tr(PGXG^{-1})=\Tr(QX),
 \label{eq:AI-Q-def}
\end{equation}
where the last two relations follow from $G\in\mathrm{SO}(N,\C)$ and
$P=P^{\mathrm T}$.

The nontrivial input is a Ward identity.
Let $E^{(ij)}$ be the antisymmetric generator with
$E^{(ij)}_{ij}=1=-E^{(ij)}_{ji}$, and set
\begin{equation}
 S(t):=\Tr\!\left(\e^{-tE^{(ij)}}Q\e^{tE^{(ij)}}X\right),
 \qquad t\in\R .
 \label{eq:AI-ward-S}
\end{equation}
Since $Q(G\e^{tE^{(ij)}})=\e^{-tE^{(ij)}}Q(G)\e^{tE^{(ij)}}$, the
deformed integrand is the original one at a translated point,
$\e^{\im S(t)}(G)=\e^{\im\Tr(QX)}(G\e^{tE^{(ij)}})$, and by the right
invariance of the Haar form \eqref{eq:AI-nu-def},
\begin{equation}
 \int_{\mathcal M_1}\dd\nu\,\e^{\im S(t)}
 =\int_{\mathcal M_1\e^{tE^{(ij)}}}\dd\nu\,\e^{\im\Tr(QX)}:
\end{equation}
the entire $t$ dependence is a shift of the integration cycle.
The shift is undone exactly as for a decaying holomorphic function of
one variable, where $\int_{\R}\dd z\,f(z)=\int_{\R+\im s}\dd z\,f(z)$:
the integrand is holomorphic and decays at the ends of the cycle, so
the shifted cycle can be moved back to $\mathcal M_1$ without changing
the integral, and $\int_{\mathcal M_1}\dd\nu\,\e^{\im S(t)}$ is
independent of $t$.
The first $t$-derivative gives an identity linear in $Q_{ij}$; the
quadratic $Q_{ij}^2$ needed below enters at second order, so we take
\begin{equation}
 \frac{\dd^{2}}{\dd t^{2}}\bigg|_{t=0}
 \int_{\mathcal M_1}\dd\nu(G)\,\e^{\im S(t)}=0 .
 \label{eq:AI-ward-setup}
\end{equation}
From $\frac{\dd}{\dd t}\bigl(\e^{-tE}Q\e^{tE}\bigr)=\e^{-tE}[Q,E]\e^{tE}$
and the symmetry $Q=Q^{\mathrm T}$, direct evaluation gives
\begin{equation}
 S'(0)=-2x_{ij}Q_{ij},
 \qquad
 S''(0)=-2x_{ij}(Q_{ii}-Q_{jj}),
 \qquad x_{ij}=x_i-x_j .
\end{equation}
Since $\partial_t^2\e^{\im S}|_{t=0}
=[\im S''(0)-S'(0)^2]\e^{\im S(0)}$, Eq.~\eqref{eq:AI-ward-setup} becomes,
after division by $-2x_{ij}^{2}$, the Ward identity
\begin{equation}
 \int_{\mathcal M_1}\dd\nu(G)\,
 \left[2Q_{ij}^2+\frac{\im(Q_{ii}-Q_{jj})}{x_{ij}}\right]
 \e^{\im\Tr(QX)}=0 .
 \label{eq:AI-orbital-Ward}
\end{equation}
The other ingredient is elementary: because $Q$ is independent of
$\bm x$ and $X$ is diagonal, each derivative acts only on the exponent,
$\partial_{x_i}\e^{\im\Tr(QX)}
=\im Q_{ii}\,\e^{\im\Tr(QX)}$, and differentiation under the integral gives
\begin{equation}
 \partial_{x_i}\mathcal I
 =\im\int_{\mathcal M_1}Q_{ii}\e^{\im\Tr(QX)}\dd\nu,
 \qquad
 \sum_i\partial_{x_i}^2\mathcal I
 =-\int_{\mathcal M_1}\sum_iQ_{ii}^2
 \e^{\im\Tr(QX)}\dd\nu .
 \label{eq:AI-orbital-derivatives}
\end{equation}
The trace $\Tr Q^2=\Tr P^2$ splits the diagonal from the off-diagonal,
$\sum_iQ_{ii}^2=\sum_ip_i^2-2\sum_{i<j}Q_{ij}^2$, and
Eq.~\eqref{eq:AI-orbital-Ward} converts the off-diagonal sum into the first
derivatives of Eq.~\eqref{eq:AI-orbital-derivatives}.
Assembling the three ingredients gives
\begin{equation}
 \left[\sum_i\partial_{x_i}^2
 +\sum_{i<j}\frac{\partial_{x_i}-\partial_{x_j}}{x_i-x_j}\right]
 \mathcal I(\bm x,\bm p)
 =-\left(\sum_i p_i^2\right)\mathcal I(\bm x,\bm p).
 \label{eq:AI-orbital-radial}
\end{equation}
For arbitrary $k$, the exact Vandermonde gauge identity is
\begin{equation}
 \Delta(\bm x)^k
 \left[\sum_i\partial_{x_i}^2
 +2k\sum_{i<j}\frac{\partial_{x_i}-\partial_{x_j}}{x_i-x_j}\right]
 \Delta(\bm x)^{-k}
 =\sum_i\partial_{x_i}^2
 -2k(k-1)\sum_{i<j}\frac1{(x_i-x_j)^2}.
 \label{eq:AI-orbital-gauge}
\end{equation}
At $k=\tfrac12$, Eqs.~\eqref{eq:AI-orbital-radial} and
\eqref{eq:AI-orbital-gauge} imply
\begin{equation}
 \mathcal H_{1/2}\Psi_{1/2}(\bm x,\bm p)
 =\left(\sum_i p_i^2\right)\Psi_{1/2}(\bm x,\bm p).
 \label{eq:AI-orbital-Calogero}
\end{equation}
Thus, the function in Eq.~\eqref{eq:AI-orbital-I} is an exact
Calogero eigenfunction.
The two prefactors play different roles: $\Delta(\bm x)^{1/2}$ is what
converts the first-order drift term of Eq.~\eqref{eq:AI-orbital-radial}
into the inverse-square potential, while the $x$-independent factor
$\Delta(-\im\bm p)^{1/2}$ passes through $\mathcal H_{1/2}$ untouched and is
instead fixed by the boundary condition of the next subsection, where it
cancels the saddle-point prefactor and normalizes the identity-channel
amplitude to unity.

\subsection{Identity-channel scattering asymptotics}
We next determine the boundary condition.
Keeping $\bm p$ fixed, we set
\begin{equation}
 \bm x=R\bm\xi,
 \qquad
 \tau_{ij}^{(0)}=-\im(p_i-p_j)(\xi_i-\xi_j),
 \qquad R\longrightarrow\infty,
 \label{eq:AI-large-gap-scaling}
\end{equation}
in a closed subsector with $\operatorname{Re}\tau_{ij}^{(0)}>0$ for every $i<j$.
This condition makes every quadratic fluctuation about $G=I$ exponentially
decaying.
The stationary equation under antisymmetric variations is $[Q,X]=0$.
For pairwise distinct $x_i$ and $p_i$, it forces $G$ to represent a Weyl
permutation; the positive-Hermitian identity component contains only $G=I$,
while the other permutations lie on different steepest-descent cycles.
Near this identity critical point, we write
$G=\e^{2\im B}$ with $B^{\mathrm T}=-B$ and $B_{ij}\in\R$.
Expansion about $B=0$ gives
\begin{equation}
 \im\Tr(PGXG^{-1})
 =\im R\bm p\cdot\bm\xi
 -4R\sum_{i<j}\tau_{ij}^{(0)}B_{ij}^2
 +O(R\|B\|^3).
 \label{eq:AI-saddle-expansion}
\end{equation}
By Eq.~\eqref{eq:AI-nu-explicit} with $j_N(0)=1$, the measure at the
identity is $\nu_0\,2^{N(N-1)/2}\prod_{i<j}\dd B_{ij}$, and the Gaussian
fluctuations give
\begin{equation}
 \mathcal I(R\bm\xi,\bm p)
 =\nu_0\frac{\pi^{N(N-1)/4}}{R^{N(N-1)/4}}
 \frac{\e^{\im R\bm p\cdot\bm\xi}}
 {[\Delta(\bm\xi)\Delta(-\im\bm p)]^{1/2}}
 [1+o(1)],
 \label{eq:AI-orbital-saddle}
\end{equation}
where we used
$\prod_{i<j}\tau_{ij}^{(0)}=\Delta(\bm\xi)\Delta(-\im\bm p)$,
with the square-root branches fixed by the domain specified above.
Since
$\Delta(R\bm\xi)^{1/2}=R^{N(N-1)/4}\Delta(\bm\xi)^{1/2}$,
the Vandermonde prefactor in Eq.~\eqref{eq:AI-orbital-I} cancels the
entire fluctuation determinant.
The invariant measure carries a single overall normalization; the choice
$\nu_0=\pi^{-N(N-1)/4}$ sets the incident plane wave to unit amplitude,
\begin{equation}
 \Psi_{1/2}(R\bm\xi,\bm p)
 =\e^{\im R\bm p\cdot\bm\xi}[1+o(1)] .
 \label{eq:AI-leading-free}
\end{equation}
At $N=2$, we verified Eq.~\eqref{eq:AI-leading-free} against the exact
Hankel solution of the End Matter.
The chosen cycle $\mathcal M_1$ passes through $G=I$ only.
Steepest-descent cycles through the other Weyl permutations would instead
produce the plane waves $\e^{\im R\sum_i p_i\xi_{\sigma(i)}}$ and hence
different scattering channels.

Equations~\eqref{eq:AI-orbital-Calogero} and
\eqref{eq:AI-leading-free} are precisely the eigenvalue equation and the
identity-channel boundary condition defining the asymptotically free
scattering state.
Both were derived under the explicit condition
$\operatorname{Re}\tau_{ij}^{(0)}>0$ for all $i<j$; holomorphic continuation extends them
throughout the collision-free domain, and neither the Calogero equation nor
the unit coefficient of the identity-channel plane wave is changed by the
continuation.

\subsection{Importance integral for the spacing}
The observable is defined in Section~\ref{sec:aii-spacing}
[Eqs.~\eqref{eq:rho-origin} and \eqref{eq:Palm-spacing}]; we evaluate
it from the positive orbit representation of
Eq.~\eqref{eq:AI-orbit-jpdf} [Section~\ref{sec:polar-ai}], in closed
form at $N=2,3$ and by Monte Carlo numerical integration at
$N=4,5,6$; no random matrices are sampled anywhere in this evaluation.
We first reduce Eq.~\eqref{eq:Palm-spacing}.
Setting $z_1=0$ removes two real variables; by permutation symmetry,
the minimizing root may be taken to be $z_2$ at the cost of a factor
$N-1$, with the spectators constrained to $|z_j|\ge s$ for $j\ge3$; the
radial delta function fixes $|z_2|=s$; and the invariance of the
integrand under a common phase $\bm z\mapsto\e^{\im\theta}\bm z$ sets
$z_2=s>0$ at the cost of $2\pi$.
With $\bm z_0:=(0,s,z_3,\ldots,z_N)$ and
$D_N(\bm z_0):=\prod_{i<j}|z_i-z_j|^2$, the product running over all
pairs of $\bm z_0$, this gives the exact
$[2(N-2)+N(N-1)/2]$-dimensional integral
\begin{equation}
 P_N^{\AId}(s)
 =\frac{(N-1)\,2\pi s\,K_N}{\rho_1(0)}\,\e^{-s^2}
 \int\limits_{|z_3|,\ldots,|z_N|\ge s}
 \e^{-\sum_{j\ge3}|z_j|^2}\,D_N(\bm z_0)
 \int j_N(B)\,\e^{-\mathcal V_B(\bm z_0)}\,\dd B
 \prod_{j\ge3}\dd^2z_j ,
 \label{eq:AI-Palm-reduced}
\end{equation}
with $\rho_1(0)$ the class-$\AId$ value of Eq.~\eqref{eq:rho-origin}.
We evaluate the integral of Eq.~\eqref{eq:AI-Palm-reduced}
numerically, by plain Monte Carlo integration, after one change of
variables that removes all of its stiffness.
Expanding Eq.~\eqref{eq:AI-orbit-potential} about $B=0$ gives
$\mathcal V_B=\sum_{i<j}(2|z_i-z_j|B_{ij})^2+O(\|B\|^3)$, so each orbit
coordinate is rescaled by its own gap,
\begin{equation}
 B_{ij}=\frac{Y_{ij}}{2|z_i-z_j|},
 \qquad
 \dd B=\prod_{i<j}\frac{\dd Y_{ij}}{2|z_i-z_j|}:
 \label{eq:AI-orbit-rescale}
\end{equation}
in the $Y$ variables, the orbit integrand
$j_N\,\e^{\|Y\|^2-\mathcal V_B}$ is $O(1)$ uniformly in the spectrum,
and the Jacobian combines with the Vandermonde as
$D_N(\bm z_0)\prod_{i<j}(2|z_i-z_j|)^{-1}
=2^{-N(N-1)/2}\prod_{i<j}|z_i-z_j|$.
Drawing $Y_{ij}\sim\mathcal N(0,\tfrac12)$ and the spectators from the
Gaussian factor restricted to $|z_j|\ge s$---sampled without rejection
as $|z_j|^2=s^2+\mathrm{Exp}(1)$ with uniform phase---reduces
Eq.~\eqref{eq:AI-Palm-reduced} to a known prefactor times the sample
mean of the $O(1)$ bracket
$\prod_{i<j}|z_i-z_j|\,j_N(B)\,\e^{\|Y\|^2-\mathcal V_B}$.
Both draws are exact and untuned, and neither carries any spectral
structure---the spectators are independent, with neither level
repulsion nor any class-dependent weight---so the entire content of
Eqs.~\eqref{eq:AI-orbit-jpdf}--\eqref{eq:AI-Palm-reduced}, including
$K_N$, $\rho_1(0)$, the factor $N-1$, $j_N$, and $\mathcal V_B$, enters
only through the integrand and is tested by the result.
At $N=6$, the largest case, $K_6=8192/(135\pi^{12})$ and
$\rho_1(0)=2/(7\pi)$; six independent runs of $2.3\times10^6$ samples
each evaluate $P_6(s)$ on the grid $s=0,0.05,\ldots,2.60$; performing
the remaining one-dimensional $r$ integrals by Simpson's rule on this
grid gives
$\int P_6(s)\dd s=1.0020(94)$ and $\langle R_6\rangle=0.8534(22)$,
the quoted errors being standard errors across the runs.
The independent matrix diagonalization described in
Section~\ref{sec:spacing} returns $\langle R_6\rangle=0.85302$ from a
computation sharing none of this machinery.

\subsection{Collision law}
The two local Calogero exponents are $k$ and $1-k$; they coalesce at
$k=\tfrac12$, so the second branch is logarithmic.
For the origin-conditioned spacing density defined in
Eq.~\eqref{eq:Palm-spacing}, direct integration gives
\begin{equation}
  P_{2}^{\AId}(s)
  =\tfrac32 s^3\e^{-s^2/2}K_0(s^2/2)
  =3s^3\log(1/s)+O(s^3),
 \label{eq:AI-collision}
\end{equation}
where $K_0$ is the modified Bessel function of the second kind.

For general $N$, the logarithm follows from an exact noncompact
invariance of the positive orbit
representation of Eq.~\eqref{eq:AI-orbit-jpdf}, whose weight is
$\mathcal V=\Tr(\widetilde Z^\dagger g\widetilde Zg^{-1})
-\Tr(\widetilde Z^\dagger\widetilde Z)$ with $g=\e^{2\im B}$.
For $u\in\mathrm{SO}(N,\C)$, the map $g\mapsto u^\dagger gu$ preserves
the set of positive Hermitian elements of $\mathrm{SO}(N,\C)$ and its
measure $j_N(B)\,\dd B$ [this is the standard group action on the
symmetric space $\mathrm{SO}(N,\C)/\mathrm{SO}(N)$, and $j_N(B)\,\dd B$ is its
invariant measure], and it replaces $\widetilde Z$ by
$u\widetilde Zu^{-1}$ in $\mathcal V$.

Consider a collision configuration: without loss of generality, by
permutation symmetry, let $z_1$ and $z_2$ be the close pair,
$|z_{12}|:=|z_1-z_2|\to0$, with the spectators $z_3,\ldots,z_N$ held
at generic positions, far from the pair and from one another.
Take $u_t=\e^{\im tE^{(12)}}$, $t\in\R$, the boost of the colliding
pair.
The maps $g\mapsto u_t^\dagger gu_t$ act on the flat integration
variable itself: they define maps $B\mapsto B_t$ of the $B$ space
[through $\e^{2\im B_t}=u_t^\dagger\e^{2\im B}u_t$] which preserve
the measure $j_N(B)\,\dd B$ and displace every point along a
noncompact orbit; at $N=2$, this is simply the translation
$\sigma\mapsto\sigma+t$ of the single flat coordinate.
We may therefore change variables on the flat $B$ space to flow
coordinates---$t$ along the orbits and transverse coordinates labeling
them---at unit Jacobian with respect to $j_N(B)\,\dd B$, so that the
measure becomes $\dd t$ times an $t$-independent transverse measure.
At $z_1=z_2$, the $(1,2)$ block of $\widetilde Z$ is a multiple of the
identity matrix, so $u_t\widetilde Zu_t^{-1}=\widetilde Z$: the
integrand is constant along every orbit, the $t$ integral is
$\int\dd t$, and the orbit integral diverges.
At small $|z_{12}|\neq0$, the $(1,2)$ block of $\widetilde Z$ deviates
from a multiple of the identity by $O(|z_{12}|)$, and the boost
amplifies the deviation exponentially: the corresponding entries of
$u_t\widetilde Zu_t^{-1}$ grow as $|z_{12}|\e^{2|t|}$, and $\mathcal V$,
quadratic in $\widetilde Z$, acquires a term of order
$|z_{12}|^2\e^{4|t|}$ [at $N=2$, exactly
$\mathcal V=|z_{12}|^2\sinh^2(2t)$].
The $t$ integral along each two-sided orbit is therefore
$\int\dd t\,\e^{-O(|z_{12}|^2\e^{4|t|})}
=\log(1/|z_{12}|)+O(1)$: the divergence is cut off at
$|t|\le t^*=\tfrac12\log(1/|z_{12}|)$, a plateau of total length
$2t^*$, exactly as in the
$N=2$ integral $\int\dd\sigma\,\e^{-s^2\sinh^2(2\sigma)}$ behind
Eq.~\eqref{eq:AI-collision}.

Since the integrand is positive and its transverse integral at fixed
$t\le t^*$ is pinned to its $t$-independent collision value by the
invariance,
\begin{equation}
 \int j_N(B)\,\e^{-\mathcal V_B}\dd B
 =C(\{z_j\}_{j\ge3})\,\log\frac1{|z_{12}|}+O(1),
 \qquad C>0 .
 \label{eq:AI-collision-generalN}
\end{equation}
With the prefactor $|\Dl|^2\sim|z_{12}|^2$ of
Eq.~\eqref{eq:AI-orbit-jpdf}, the density vanishes as
$|z_{12}|^2\log(1/|z_{12}|)$, and integration over the spectators,
whose weight is positive, preserves the positivity of coefficient:
\begin{equation}
 P_N^{\AId}(s)=C_Ns^3\log(1/s)+O(s^3),
 \qquad C_N>0\quad\text{for all }N\ge2,
 \label{eq:AI-collision-law}
\end{equation}
extending Eq.~\eqref{eq:AI-collision} to every matrix size.

\section{Finite-\texorpdfstring{$N$}{N} Palm and large-\texorpdfstring{$N$}{N} bulk checks}
\label{sec:spacing}

The exact finite-$N$ distributions derived above are reduced Palm
distributions at the origin.  One labelled eigenvalue is conditioned to lie at
$z_1=0$, and the observable is its distance
$R_{\rm nn}=\min_{j\ge2}|z_j|$ to the nearest remaining eigenvalue, as defined
in Eq.~\eqref{eq:Palm-spacing}.  This is not the global finite-$N$
nearest-neighbor distribution obtained by marking every eigenvalue and using
$\min_{j\ne i}|z_i-z_j|$.  At finite $N$ the spectral density varies across
the plane, so the global distribution mixes bulk and edge spacings and depends
on an unfolding prescription.  The origin-conditioned distribution instead
follows directly from the exact joint density and requires no finite-$N$
unfolding.

A matrix drawn from a continuous Gaussian ensemble has probability zero to
possess an eigenvalue exactly at the origin.  We therefore do not set or shift
an eigenvalue to zero in the direct-diagonalization check.  For every finite-$N$
data set, we diagonalize $5\times10^6$ independent matrices, order the distinct
eigenvalue moduli as $r_{(1)}\le r_{(2)}\le\cdots$, retain matrices with
$r_{(1)}<\epsilon$, and record $r_{(2)}$.  In the limit $\epsilon\to0$, this
finite-window conditional distribution approaches the origin Palm
distribution.  We use $\epsilon=0.10$ in the figure and verify stability
against $\epsilon=0.15$; no eigenvalue is shifted.  After rescaling each data
set to unit mean, the bars are densities in 26 equal bins on
$0\le s\le2.6$, with binomial standard errors.

For the $N=128$ bulk comparison we instead compute the ordinary complex
nearest-neighbor spacing directly.  For every eigenvalue in the central disk
$|z_i|<0.6R$, with $R=\sqrt{N/2}$ for class $\AId$ and $R=\sqrt N$ for class
$\AIId$, we evaluate
$d_i=\min_{j\ne i}|z_i-z_j|$ against the full spectrum.
For class $\AIId$, one eigenvalue is retained from each Kramers pair before
forming the distinct-level spacings.  The data comprise 5000 class-$\AId$
matrices and 2000 class-$\AIId$ matrices, giving 227490 and 92694 marked bulk
spacings, respectively.  Each class is normalized by its empirical bulk mean,
and the plotted uncertainties are matrix-cluster standard errors.

\begin{table}[p]
\caption{The coefficients of the $N=7$ survival polynomial
$H_7(u)=\sum_{k=0}^{42}h_ku^{k}$ of Eq.~\eqref{eq:aii-n7-survival}.}
\label{tab:H7}
\scriptsize
\renewcommand{\tfrac}[2]{#1/#2}
\centering
\begin{tabular}{cl}
\hline\hline
 $k$ & $h_k$\\
\hline
 $0$ & $1$\\
 $1$ & $12$\\
 $2$ & $\tfrac{2363}{33}$\\
 $3$ & $\tfrac{84130}{297}$\\
 $4$ & $\tfrac{26058622}{31185}$\\
 $5$ & $\tfrac{4075600}{2079}$\\
 $6$ & $\tfrac{20788017461}{5457375}$\\
 $7$ & $\tfrac{20642015917}{3274425}$\\
 $8$ & $\tfrac{4277023397}{471625}$\\
 $9$ & $\tfrac{263975808896}{22920975}$\\
 $10$ & $\tfrac{8087899725676}{618866325}$\\
 $11$ & $\tfrac{6831874122323746}{510564718125}$\\
 $12$ & $\tfrac{6361773221898124}{510564718125}$\\
 $13$ & $\tfrac{258248218003382}{24312605625}$\\
 $14$ & $\tfrac{12766754953764002}{1531694154375}$\\
 $15$ & $\tfrac{52938996573817}{8752538025}$\\
 $16$ & $\tfrac{655331855440439666}{160827886209375}$\\
 $17$ & $\tfrac{276916533645676}{108301606875}$\\
 $18$ & $\tfrac{1205087004548853826}{804139431046875}$\\
 $19$ & $\tfrac{1983689501685776812}{2412418293140625}$\\
 $20$ & $\tfrac{3062738911792378472}{7237254879421875}$\\
 $21$ & $\tfrac{1480666811195325364}{7237254879421875}$\\
\hline\hline
\end{tabular}\hspace{2.5em}
\begin{tabular}{cl}
\hline\hline
 $k$ & $h_k$\\
\hline
 $22$ & $\tfrac{14136674805137338856}{151982352467859375}$\\
 $23$ & $\tfrac{4320631379276130766}{108558823191328125}$\\
 $24$ & $\tfrac{12184475667716217608}{759911762339296875}$\\
 $25$ & $\tfrac{4621837501145542796}{759911762339296875}$\\
 $26$ & $\tfrac{18000939440766004}{8289946498246875}$\\
 $27$ & $\tfrac{1662070939463427748}{2279735287017890625}$\\
 $28$ & $\tfrac{1572353565656052248}{6839205861053671875}$\\
 $29$ & $\tfrac{464795062868187736}{6839205861053671875}$\\
 $30$ & $\tfrac{25692382194532712}{1367841172210734375}$\\
 $31$ & $\tfrac{99243884326328384}{20517617583161015625}$\\
 $32$ & $\tfrac{23704106312030896}{20517617583161015625}$\\
 $33$ & $\tfrac{5219754674693632}{20517617583161015625}$\\
 $34$ & $\tfrac{150193321765184}{2931088226165859375}$\\
 $35$ & $\tfrac{191667853178432}{20517617583161015625}$\\
 $36$ & $\tfrac{10393498079296}{6839205861053671875}$\\
 $37$ & $\tfrac{4438606908928}{20517617583161015625}$\\
 $38$ & $\tfrac{5431538944}{207248662456171875}$\\
 $39$ & $\tfrac{35819264}{13816577497078125}$\\
 $40$ & $\tfrac{17406208}{88820855338359375}$\\
 $41$ & $\tfrac{59392}{5921390355890625}$\\
 $42$ & $\tfrac{512}{1973796785296875}$\\
 & \\
\hline\hline
\end{tabular}
\end{table}

\begin{table}[p]
\caption{The coefficients of the $N=8$ survival polynomial
$H_8(u)=\sum_{k=0}^{56}h_ku^{k}$ of Eq.~\eqref{eq:aii-n8-survival}.}
\label{tab:H8}
\scriptsize
\renewcommand{\tfrac}[2]{#1/#2}
\centering
\begin{tabular}{cl}
\hline\hline
 $k$ & $h_k$\\
\hline
 $0$ & $1$\\
 $1$ & $14$\\
 $2$ & $\tfrac{1269}{13}$\\
 $3$ & $\tfrac{64628}{143}$\\
 $4$ & $\tfrac{18103834}{11583}$\\
 $5$ & $\tfrac{349159960}{81081}$\\
 $6$ & $\tfrac{1330476047}{135135}$\\
 $7$ & $\tfrac{123900559964}{6449625}$\\
 $8$ & $\tfrac{16215598304239}{496621125}$\\
 $9$ & $\tfrac{1161278740192}{23648625}$\\
 $10$ & $\tfrac{887018459066318}{13408770375}$\\
 $11$ & $\tfrac{463299930058312}{5746615875}$\\
 $12$ & $\tfrac{1784447759423927854}{19912024006875}$\\
 $13$ & $\tfrac{23679632542646391919}{258856312089375}$\\
 $14$ & $\tfrac{4465051324203624337}{51771262417875}$\\
 $15$ & $\tfrac{410274827300047506553}{5435982553876875}$\\
 $16$ & $\tfrac{334675185398457218074}{5435982553876875}$\\
 $17$ & $\tfrac{1277153908717580711152}{27179912769384375}$\\
 $18$ & $\tfrac{304988554604467805882}{9059970923128125}$\\
 $19$ & $\tfrac{1851143768637894495811}{81539738308153125}$\\
 $20$ & $\tfrac{504922556080155820699}{34945602132065625}$\\
 $21$ & $\tfrac{31914698569432852736866}{3669288223866890625}$\\
 $22$ & $\tfrac{934627446157829814630511}{188356795491833718750}$\\
 $23$ & $\tfrac{69026208384819435144739}{25685017567068234375}$\\
 $24$ & $\tfrac{1173007296112672218478318}{847605579713251734375}$\\
 $25$ & $\tfrac{20913265226747889790871}{30822021080481881250}$\\
 $26$ & $\tfrac{4479684298414519059363479}{14126759661887528906250}$\\
 $27$ & $\tfrac{1797885859171603847862277}{12714083695698776015625}$\\
 $28$ & $\tfrac{1276088465140014473074804}{21190139492831293359375}$\\
\hline\hline
\end{tabular}\hspace{2em}
\begin{tabular}{cl}
\hline\hline
 $k$ & $h_k$\\
\hline
 $29$ & $\tfrac{118976148330631498054144}{4854468320175896296875}$\\
 $30$ & $\tfrac{38196638100993759216802664}{4004936364145114444921875}$\\
 $31$ & $\tfrac{6462900130108869437750932}{1820425620065961111328125}$\\
 $32$ & $\tfrac{460366467828666309268922}{364085124013192222265625}$\\
 $33$ & $\tfrac{319601802248216776966109}{741654882249095267578125}$\\
 $34$ & $\tfrac{1688333488520116915079314}{12014809092435343334765625}$\\
 $35$ & $\tfrac{877724311380138204751672}{20024681820725572224609375}$\\
 $36$ & $\tfrac{25890058516847452587896}{1980463037214617033203125}$\\
 $37$ & $\tfrac{671402454849411214686268}{180222136386530150021484375}$\\
 $38$ & $\tfrac{182655545434149640053034}{180222136386530150021484375}$\\
 $39$ & $\tfrac{142144962442641475469456}{540666409159590450064453125}$\\
 $40$ & $\tfrac{1671715122706116023476}{25746019483790021431640625}$\\
 $41$ & $\tfrac{392341125452314538216}{25746019483790021431640625}$\\
 $42$ & $\tfrac{5499657163083613200392}{1621999227478771350193359375}$\\
 $43$ & $\tfrac{1156723092098219366416}{1621999227478771350193359375}$\\
 $44$ & $\tfrac{45829356121352657704}{324399845495754270038671875}$\\
 $45$ & $\tfrac{3869361377273973904}{147454475225342850017578125}$\\
 $46$ & $\tfrac{7371136152115154096}{1621999227478771350193359375}$\\
 $47$ & $\tfrac{236352260483616608}{324399845495754270038671875}$\\
 $48$ & $\tfrac{11587364605508288}{108133281831918090012890625}$\\
 $49$ & $\tfrac{1103720280077696}{77238058451370064294921875}$\\
 $50$ & $\tfrac{2759536843742528}{1621999227478771350193359375}$\\
 $51$ & $\tfrac{3157065249664}{17824167334931553298828125}$\\
 $52$ & $\tfrac{59227096576}{3780883980136996154296875}$\\
 $53$ & $\tfrac{1421777152}{1260294660045665384765625}$\\
 $54$ & $\tfrac{4774144}{77160897553816248046875}$\\
 $55$ & $\tfrac{11776}{5144059836921083203125}$\\
 $56$ & $\tfrac{512}{12002806286149194140625}$\\
 & \\
\hline\hline
\end{tabular}
\end{table}

\section{Reflection matrices of disordered conductors}
\label{sec:transport}

This section records the numerical procedure behind the reflection-matrix
spacings of the End Matter; the lattice models, parameters, and the
transposition symmetry of $r$ are given there.
The leads are clean semi-infinite cubic lattices attached to the two $x$
surfaces, with hopping $t_{\rm L}=1$, zero onsite potential, no spin-orbit
coupling, and the same transverse periodic boundaries as the sample; each
surface site couples to the corresponding lead site by $-t_{\rm L}$ (times
$\sigma_0$ in class AII).

The scattering matrix $S(E)$ is obtained by Kwant~\cite{Groth2014}.
Numerical lead modes carry arbitrary phases and degenerate-subspace bases;
before diagonalizing $r$, we therefore rotate the incoming and outgoing modes
to a common real transverse eigenmode basis $X$,
\begin{equation}
U_{\rm in/out}=X^\dagger\Phi_{\rm in/out},
\qquad r=U_{\rm out}r_{\rm Kwant}U_{\rm in}^\dagger,
\label{eq:sm-reflection-basis}
\end{equation}
with $\Phi_{\rm in/out}$ the velocity-normalized Kwant mode matrices.

For each distinct eigenvalue $z_i$, the raw spacing is
$d_i=\min_{j\ne i}|z_i-z_j|$, with midpoint $m_i$; the central $30\%$ of the
spectrum by modulus is retained.
The ensemble-averaged planar density is estimated by Gaussian kernel
smoothing~\cite{Xiao2022},
\begin{equation}
\widehat\rho_\sigma(z)=\frac{1}{N_{\rm s}}\sum_{a=1}^{N_{\rm s}}\sum_k
\frac{\exp[-|z-z_{ak}|^2/(2\sigma^2)]}{2\pi\sigma^2},
\qquad \sigma=3\bar d,
\label{eq:sm-reflection-density}
\end{equation}
with $\bar d$ the mean raw spacing in the retained bulk, and the plotted
spacing is locally unfolded to unit mean,
\begin{equation}
s_i=\frac{d_i\sqrt{\widehat\rho_\sigma(m_i)}}
{\left\langle d\sqrt{\widehat\rho_\sigma(m)}\right\rangle_{\rm bulk}}.
\label{eq:sm-reflection-unfolding}
\end{equation}
The figure of the End Matter uses $400$ independent realizations per class;
its error bars are standard errors of the per-realization histograms.

\bibliography{references}